\documentclass[nofootinbib,aps,10pt,twocolumn]{revtex4-1}

\usepackage{array,amsmath,amsthm}
\usepackage{mathtools}
\usepackage{graphicx}
\usepackage{color}
\usepackage{amssymb}

\begin{document}
\title{Leptobaryons as Majorana Dark Matter}
\author{Sebastian Ohmer}
\email{sebastian.ohmer@mpi-hd.mpg.de}
\affiliation{Particle and Astro-Particle Physics Division \\
Max-Planck Institut fuer Kernphysik {\rm{(MPIK)}} \\
Saupfercheckweg 1, 69117 Heidelberg, Germany}
\author{Hiren H. Patel}
\email{hiren.patel@mpi-hd.mpg.de}
\affiliation{Particle and Astro-Particle Physics Division \\
Max-Planck Institut fuer Kernphysik {\rm{(MPIK)}} \\
Saupfercheckweg 1, 69117 Heidelberg, Germany}
\begin{abstract}
We explore the dark matter and collider phenomenology of the minimal gauged $\text{U(1)}_\text{B}$ model, consisting of a leptophobic $Z_\text{B}$ gauge boson, and an accompanying Higgs $S_\text{B}$.  By requirement of anomaly cancellation, the fermion sector naturally contains a dark matter candidate---a Majorana isosinglet $\chi$ stabilized by an inherent $Z_2$ symmetry.  The absence of evidence for $Z$ prime dijet resonances at the LHC suggests that the scale of symmetry breaking is $\Lambda_{\text{B}}\gtrsim 500$ GeV.  Saturation of dark matter abundance together with limits on the direct detection cross section (dominated by Higgs exchange) constrains the Higgs mixing angle to $|\theta|\lesssim 0.22$.  For small mixing angles of $|\theta| \lesssim 10^{-3}$, the $\mathcal{O}(10\%)$ branching fractions of the fermion loop-mediated $S_\text{B}\rightarrow\gamma \gamma$, $Z\gamma$, $ZZ$ modes may provide clues about the fermion content of the model at the LHC.
\end{abstract}
\maketitle
\section{Introduction}
{\sc Classical} equations of motion derived from the standard model Lagrangian imply exact conservation of the baryonic and leptonic currents
\begin{equation*}
J_\text{B}^\mu = \sum_{q} {\textstyle  \frac{1}{3}} \bar{q} \gamma^\mu q\,, \qquad
J_\text{L}^\mu = \sum_{\mathclap{i={e,\mu,\tau}}}( \bar\ell_i \gamma^\mu \ell_i + \bar\nu_i \gamma^\mu \nu_i)\,.
\end{equation*}
In contrast with other conserved currents such as electromagnetic, color, and weak neutral/charged currents, these currents do not couple to any known gauge fields.  This observation has motivated a number of authors to speculate on the possibility that these currents separately source fields that may have so far escaped detection \mbox{\cite{FileviezPerez:2010gw,Dulaney:2010dj,FileviezPerez:2011pt,Duerr:2013dza,Perez:2014qfa}}.  It is recognized, however, that these symmetries are quantum mechanically anomalous, reflected by the nonconservation of the corresponding currents 
\begin{equation*}
\partial_\mu J^\mu_\text{B} = \partial_\mu J^\mu_\text{L} = \frac{g^2}{16\pi^2}\frac{3}{2} \vec{\widetilde{W}}_{\mu\nu}\!\cdot\!\vec{W}^{\mu\nu} - \frac{g'^2}{16\pi^2}\frac{3}{2} \widetilde{B}_{\mu\nu}B^{\mu\nu}\,,
\end{equation*}
which makes them unsuitable to serve as sources for new gauge fields. To alleviate this difficulty, the standard model baryonic and leptonic currents are each extended by introducing additional fermionic degrees of freedom with opposite anomalous contributions, such that the totals are anomaly free:
\begin{gather}
\nonumber \begin{aligned}
J^\mu_\text{B,tot} &= J^\mu_\text{B} + J_\text{B,ext}^\mu\\ 
J^\mu_\text{L,tot} &= J^\mu_\text{L} + J_\text{L,ext}^\mu\,
\end{aligned}\\
\label{eq:AnomalyCondition1}\partial_\mu J^\mu_\text{B,tot} = \partial_\mu J^\mu_\text{L,tot} = 0\,.
\end{gather}
The form of the extra pieces $J_\text{B,ext}^\mu$ and $J_\text{L,ext}^\mu$, that are built out of the new fermions, are model dependent.  It should also be checked that the modified weak isospin and hypercharge currents remain anomaly free upon the addition of new degrees of freedom
\begin{equation}\label{eq:AnomalyCondition2}
\partial_\mu J^{\mu}_\text{SU(2)} = \partial_\mu J^{\mu}_\text{U(1)} = 0\,.
\end{equation}
The total currents $J^\mu_\text{B,tot}$, $J^\mu_\text{L,tot}$ can then be consistently coupled to gauge fields $Z_\text{B}$ and $Z_\text{L}$, thereby enlarging the particle physics symmetry gauge group to
\begin{equation}\label{eq:gaugeGroup}
\text{SU(3)}_\text{C} \otimes \text{SU(2)}_L \otimes \text{U(1)}_Y \otimes \text{U(1)}_\text{B} \otimes \text{U(1)}_\text{L}.
\end{equation}

This line of thought has lead to interesting physical results such as an understanding of the stability of dark matter (without imposing an \emph{ad hoc} discrete symmetry) \cite{FileviezPerez:2010gw}, and possible unification among all group factors in (\ref{eq:gaugeGroup}) at scales much lower than the traditional grand unification scale without rapid proton decay \cite{Perez:2014kfa}.

The earliest models satisfying (\ref{eq:AnomalyCondition1}) and (\ref{eq:AnomalyCondition2}) contained charged fermions that significantly modified the nominal properties 
of the standard model Higgs boson, and were subsequently ruled out by recent experimental findings at the LHC (see \cite{Perez:2013tea} for a brief review).  

Currently, there are two closely related viable minimal models.  They contain a new set of colorless fermionic degrees of freedom, called \emph{leptobaryons}, carrying both baryon and lepton numbers as required by anomaly cancellation.  The model in \cite{Duerr:2013dza} was constructed first, and its phenomenology was studied in \cite{Perez:2013tea, Duerr:2014wra} (see Appendix \ref{sec: model VA} for its formulation). In this model, the extended baryonic and leptonic currents contain both vector and axial-vector parts
\begin{equation}
J_\text{B,ext}^\mu = J_\text{L,ext}^\mu = J_\text{V}^\mu + J_\text{A}^\mu\,.
\end{equation}
The axial charges are fixed by requirement of anomaly cancellation, but the overall normalization of vector charges is unrestricted and represent a free parameter of the theory.  Throughout this paper, we refer to this extension as \emph{leptobaryon model VA} after the form of its $\text{U(1)}_\text{B,L}$ couplings.

Recently \cite{Perez:2014qfa}, a more systematic study revealed a second consistent model which contains fewer fermion multiplets.  In this model, the extended baryonic and leptonic currents have no vector part; so we call this extension \emph{leptobaryon model A}.  So far, there has been no study of its phenomenology.  Our goal for the present paper is to initiate the exploration of dark matter and collider phenomenology.

Throughout this article we only investigate the phenomenology of low scale $\text{U(1)}_\text{B}$ breaking, and neglect $\text{U(1)}_\text{L}$.  This choice may be viewed within the leptobaryon model as taking the high-scale limit of lepton number breaking.  However, there is no compelling reason not to consider $\text{U(1)}_\text{L}$ breaking near the electroweak scale; in fact, its phenomenology has been studied recently in the context of leptobaryon model VA \cite{Aranda:2014zta, Schwaller:2013hqa}.  However, we leave the corresponding exploration in leptobaryon model A for the future.

The rest of the paper is organized as follows.  In Section \ref{sec: model A} we formulate the model and analyze the ground state particle spectrum, and in Section \ref{sec: Constraints} we discuss phenomenological constraints on the model including vacuum stability. In Section \ref{sec:Baryogenesis}, we make a brief comment on baryogenesis in this model.  The longer Sections \ref{sec:DarkMatter} and \ref{sec: Colliders} are on dark matter and collider phenomenology respectively. We draw conclusions in section \ref{sec: Conclusions}.

\section{Formulation of leptobaryon model A}
\label{sec: model A}

\subsection{Field content and Lagrangian}
\begin{table}
\begin{tabular}{ccccc}
\hline\hline
 & $\text{SU(2)}_L$ & $\text{U(1)}_Y$ & $\text{U(1)}_\text{B}$ & $\text{U(1)}_\text{L}$\\
 Gauge fields & $\vec{W}^\mu$ & $B^\mu$ & $Z_\text{B}^\mu$ & $(Z_\text{L}^\mu)$\\[1mm]
\hline Fermions\\

$\qquad(\overline{\nu}_{e,\mu,\tau})$ & $\mathbf{1}$ & $0$ & $0$ & $-1$ \\
$\Psi$ & $\mathbf{2}$ & $\phantom{+}1/2$ & $\phantom{+}3/2$ & $\phantom{+}3/2$\\
$\overline{\Psi}$ & $\overline{\mathbf{2}}$ & $-1/2$ & $\phantom{+}3/2$ & $\phantom{+}3/2$\\
$\vec{\Sigma}$ & $\mathbf{3}$ & $0$ & $-3/2$ & $-3/2$\\
$\chi$ & $\mathbf{1}$ & $0$ & $-3/2$ & $-3/2$
\\[1mm] \hline
Scalar fields\\
$H$ & $\mathbf{2}$ & $\phantom{+}1/2$ & $0$ & $0$ \\
$S_\text{B}$ & $\mathbf{1}$ & $0$ & $3$ & $3$ \\
$(S_\text{L})$ & $\mathbf{1}$ & $0$ & $0$ & $2$ \\
\hline\hline
\end{tabular}
\caption{Field content of leptobaryon model A, with the fermions given in two-component left-handed notation.  All fields are color singlets, and the hypercharge is normalized according to $Q = T^3 + Y$.  Leptonic fields bracketed in parenthesis $Z_\text{L}^\mu$, $S_\text{L}$, and $\bar\nu_{e,\mu,\tau}$ are not considered for the present phenomenological study.}
\label{tab:contentModelA}
\end{table}
The gauge-Higgs sector of the standard model is augmented by the addition of a $\text{U(1)}_\text{B}$ gauge boson $Z_\text{B}$ and a Higgs singlet $S_\text{B}$
\begin{multline}
\mathcal{L} = \textstyle -\frac{1}{4}Z_{\text{B}\mu\nu}Z_\text{B}^{\mu\nu} - \frac{1}{2}\sin (\epsilon) B_{\mu\nu}Z_\text{B}^{\mu\nu}\\
+(D_\mu H)^\dag D^\mu H + (D_\mu S_\text{B})^* D^\mu S_\text{B} - V(H,S_\text{B})\,,
\end{multline}
where $\sin(\epsilon)$ parametrizes the gauge kinetic mixing with the hypercharge gauge field $B_\mu$.
The covariant derivatives acting on the scalar fields are
\begin{gather}
\begin{aligned}
 D_\mu H &= \textstyle (\partial_\mu + ig\vec{T}\!\cdot\!\vec{W}_\mu + \frac{1}{2}ig'B_\mu)H\,,\\
 D_\mu S_\text{B} &= \textstyle (\partial_\mu + 3ig_\text{B}Z_{\text{B}\mu})S_\text{B}\,,
\end{aligned}
\end{gather}
and the tree-level scalar potential is
\begin{multline}\label{eq:modelpotential}
V(H,S_\text{B}) = -\mu^2 H^\dag H + \lambda (H^\dag H)^2 - \mu_S^2 S_\text{B}^* S_\text{B} \\
+ b(S_\text{B}^* S_\text{B})^2 + a H^\dag H S_\text{B}^* S_\text{B}\,.
\end{multline}
Four Weyl fermion multiplets, the leptobaryons, are introduced and their charge assignments (see Table \ref{tab:contentModelA}) are fixed by anomaly cancellation.  The scalar field $S_\text{B}$ is assigned a baryonic charge of +3 to allow for Yukawa-type couplings with the leptobaryons, which in two-component spinor notation are given as
\begin{equation}\label{eq:YukawaWithSB}
\mathcal{L} = - y_\psi S^*_\text{B} \overline{\Psi}\Psi - \frac{y_\Sigma}{2} S_\text{B} \vec\Sigma\cdot\vec\Sigma - \frac{y_\chi}{2} S_\text{B} \chi \chi + \text{c.c.}\,,
\end{equation}
and would generate their masses through the Higgs mechanism.  In general, the standard model Higgs doublet can also couple to the leptobaryons
\begin{multline}\label{eq:YukawaWithH}
\mathcal{L} = -\lambda_1 (H \overline\Psi)\chi - \lambda_2 (H^\dag \Psi) \chi\\
 - \lambda_3 H^\dag (\vec{T}\!\cdot\!\vec\Sigma)\Psi -\lambda_4 \overline\Psi (\vec{T}\!\cdot\!\vec\Sigma) H + \text{c.c.}\,.
\end{multline}
These interactions mediate the decay of the heavier leptobaryons into lighter ones via the emission of Higgs quanta.

It is useful to note that the pure $S_\text{B}$ Yukawa interactions in (\ref{eq:YukawaWithSB}) exhibit reflection symmetries corresponding to the transformations
\begin{gather}\label{eq:LeptobaryonDiscreteSymmetry}
\begin{aligned}
(Z_2)_\Psi&: \enspace\Psi \rightarrow -\Psi \text{ and } \overline\Psi \rightarrow -\overline\Psi\\
(Z_2)_\Sigma&: \enspace\vec\Sigma \rightarrow -\vec\Sigma\\
(Z_2)_\chi &: \enspace \chi\rightarrow -\chi\,.
\end{aligned}
\end{gather}
The addition of standard model Higgs Yukawa interactions in (\ref{eq:YukawaWithH}) explicitly breaks this symmetry leaving a single accidental $Z_2$ symmetry corresponding to the combined reflection of all leptobaryons, and preserves the stability of the lightest leptobaryon.  These interactions contribute to custodial isospin violation, and can also provide a source of $CP$ violation. 

\subsection{Electroweak/baryonic vacuum structure and particle spectrum}\label{sec:ParticleSpectrum}
Phenomenological viability of the model requires that both the standard model Higgs doublet and the baryonic Higgs singlet acquire vacuum expectation values ({\sc vev}s) in the ground state
\begin{gather}
\begin{aligned}
H^\top &= \textstyle\big(\phi^\pm,\,\frac{1}{\sqrt 2}(v + h' + i\phi^0)\big)\\
S_\text{B} &= \textstyle \frac{1}{\sqrt{2}}(v_\text{B} + s' + i \phi_\text{B})\,,
\end{aligned}
\end{gather}
where $v=246$ GeV, and $v_\text{B}$ is the {\sc vev} of the baryonic Higgs conventionally taken to be positive by the $Z_2$ symmetry of the potential.

The Higgs portal coupling $-a H^\dag H S_\text{B}^* S_\text{B}$ induces mixing between the neutral components of the doublet and singlet Higgs fields.  Our convention (opposite to that in \cite{Duerr:2014wra}) for the transformation to mass eigenstates is
\begin{equation}\label{eq:scalarRotation}
\begin{pmatrix}h' \\ s' \end{pmatrix} = \begin{pmatrix}\cos\theta & \sin\theta \\ -\sin\theta & \cos\theta\end{pmatrix}\begin{pmatrix}h\\s\end{pmatrix},\hspace{3mm} \textstyle -\frac{\pi}{4} < \theta < \frac{\pi}{4}\,.
\end{equation}
It is a straightforward exercise to apply the minimization and local stability conditions to exchange the five potential parameters $\mu^2,\,\mu_S^2,\,\lambda,\,b,\,a$ in terms of the {\sc vev}s $v$ and $v_\text{B}$, masses $m_H$ and $m_S$, and the mixing angle $\theta$ (see Appendix \ref{sec: ScalarPotentialParameter}).
Global tree-level vacuum stability (boundedness from below) requires that $\lambda >0$, $b>0$ and $4\lambda b - a^2 > 0$, which are automatically satisfied by taking positive {\sc vev}s and masses. 

\begin{figure}
\includegraphics[width=0.8\columnwidth]{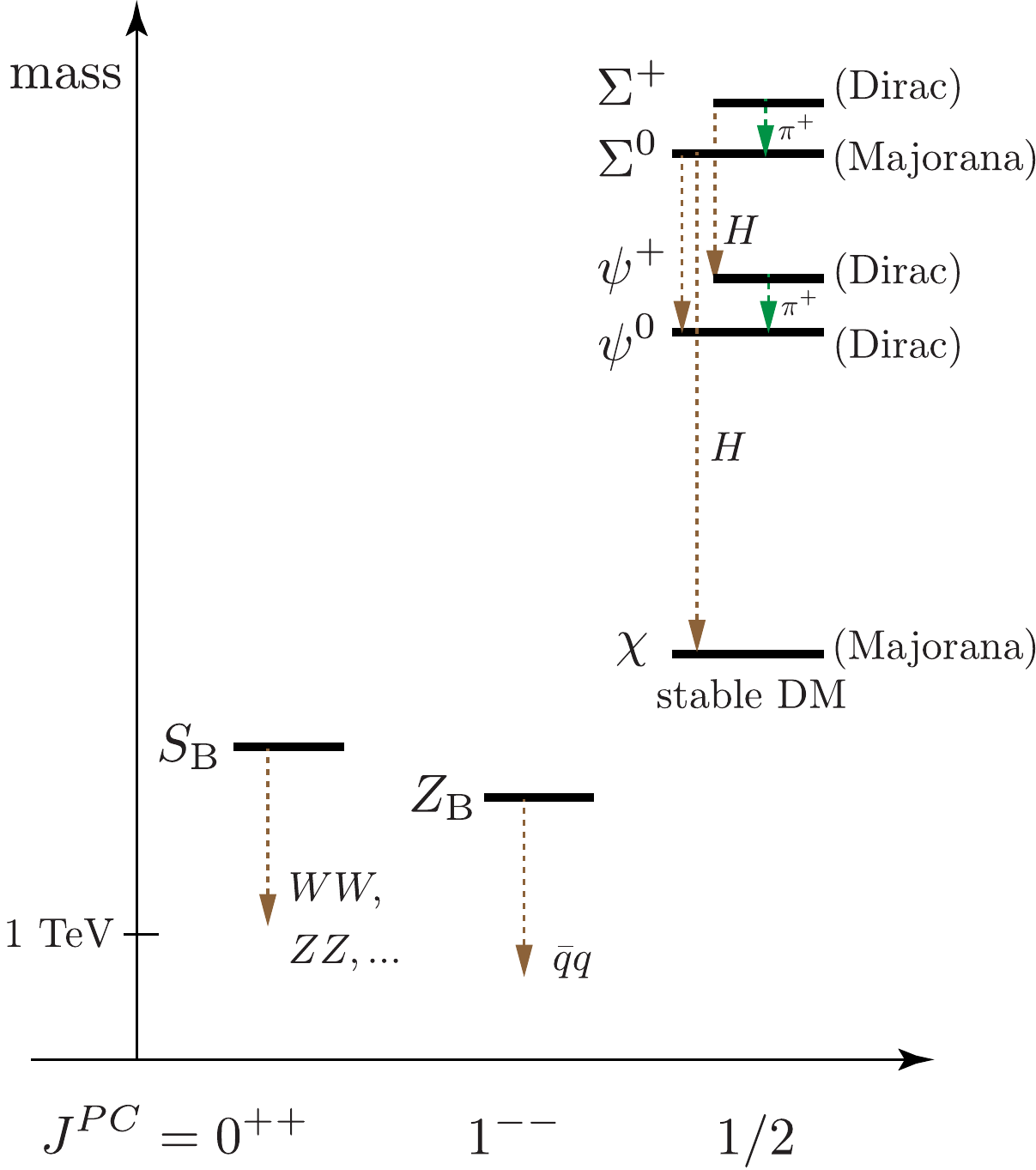}
\caption{An illustration of the mass spectrum of leptobaryon model A considered in this study. We take the isosinglet $\chi$ as the lightest leptobaryon, which is automatically stable by the $Z_2$ symmetry in (\ref{eq:LeptobaryonDiscreteSymmetry}).
}
\label{fig:LeptobaryonSpectrum}
\end{figure}

Upon spontaneous symmetry breaking of $U(1)_\text{B}$, the baryonic gauge field $Z_\text{B}$ acquires a mass given by 
\begin{equation}
m_{Z_\text{B}} = 3 g_\text{B} v_\text{B}\,,
\end{equation}
and the Yukawa interactions in (\ref{eq:YukawaWithSB}) lead to masses of the leptobaryons $\Psi$, $\Sigma$ and $\chi$ given by
\begin{equation}
m_{\psi,\Sigma,\chi} = \frac{y_{\psi,\Sigma,\chi}}{\sqrt{2}}\,v_\text{B}\,.
\end{equation}
The Weyl degrees of freedom are organized such that when expressed in terms of eigenstates of the charge operator, both members of the isodoublet $\Psi = (\psi^+, \psi^0)$ and the charged component of the isotriplet $\Sigma^+$ are Dirac fermions, and the neutral member of the isotriplet $\Sigma^0$ and the isosinglet $\chi$ are Majorana fermions.  Note that the leptobaryons carry the same electroweak quantum numbers as the gauginos and Higgsinos in the minimal supersymmetric standard model.

In what follows, we take the Majorana fermion $\chi$ to be the lightest leptobaryon (stabilized by an inherent $Z_2$ symmetry) and therefore as a dark matter candidate.  Electroweak self-energies induce mass splittings between the charged and neutral components of the isomultiplets given by\footnote{Here, we cite the values given in \cite{Cirelli:2005uq}.}:
\begin{gather}
\begin{aligned}
m_{\psi^+} - m_{\psi^0} &\\
& \hspace{-10mm} = \textstyle\frac{1}{2}\alpha\, m_Z \approx \text{341 MeV} \\
m_{\Sigma^\pm}-m_{\Sigma^0} & \\
& \hspace{-10mm} = \textstyle\frac{1}{2}\big[\alpha\, m_Z - \alpha_\text{w}(m_Z-m_W)\big] \approx\text{166 MeV}\,.
\end{aligned}  
\end{gather}
This allows the charged leptobaryons to decay to their respective neutral partners via a charged pion emission.  The resulting particle spectrum is sketched in Fig. \ref{fig:LeptobaryonSpectrum}.

Finally, the Higgs interactions with the leptobaryons in (\ref{eq:YukawaWithH}) lead to mixing among the multiplets, and mediate the decay of all leptobaryons to the lightest one via Higgs boson emission.  We shall simplify the phenomenology by taking the couplings $\lambda_{1}\ldots\lambda_4$ to be small\footnote{Note that this assumption is compatible with technical naturalness because the symmetry in (\ref{eq:LeptobaryonDiscreteSymmetry}) is recovered in the limit where these couplings vanish.} even though this is not phenomenologically necessary. 

\subsection{Electroweak and extended baryonic currents}
After investigating the scalar interactions of the leptobaryons we now turn to the gauge interactions. In our notation, the gauge interaction Lagrangian of the leptobaryons is given by 
\begin{multline*}
\mathcal{L} = -e A_\mu J^\mu _\text{EM} - \textstyle \frac{g}{c_\text{w}} Z_\mu J^\mu_\text{NC}\\
 - g \big(W_\mu^+ J^{-\mu}_\text{CC} + \text{c.c.}\big) - g_\text{B} Z_{\text{B}\mu} J^\mu_\text{B}\,,
\end{multline*}
where in four-component notation, the extended electroweak currents are
\begin{align}\label{eq:extEWCurrentModelA}
\nonumber J_\text{EM}^\mu &= \overline{\psi^+} \gamma^\mu \psi^+ + \overline{\Sigma^+} \gamma^\mu \Sigma^+\\
\nonumber J_\text{NC}^\mu &= (\textstyle\frac{1}{2}-s_\text{w}^2)\overline{\psi^+} \gamma^\mu \psi^+ -\frac{1}{2} \overline{\psi^0}\gamma^\mu\psi^0 + c_\text{w}^2\overline{\Sigma^+}\gamma^\mu\Sigma^+\\
J_\text{CC}^{-\mu} &= \textstyle \frac{1}{\sqrt{2}} \overline{\psi^+}\gamma^\mu\psi^0 - \overline{\Sigma^+}\gamma^\mu\Sigma^0\,,
\end{align}
and the extended baryonic (axial-vector) current is
\begin{multline}\label{eq:extBaryonCurrentModelA}
J_\text{B,ext}^\mu  = -\textstyle\frac{3}{2}\big(\overline{\psi^+}\gamma^\mu\gamma_5\psi^+ + \overline{\psi^0}\gamma^\mu\gamma_5\psi^0\big)\\
\textstyle+\frac{3}{2}\big(\overline{\Sigma^+}\gamma^\mu\gamma_5\Sigma^++\frac{1}{2}\overline{\Sigma^0}\gamma^\mu\gamma_5\Sigma^0\big)+\frac{3}{4}\bar\chi\gamma^\mu\gamma_5\chi\,.
\end{multline}
The couplings of leptobaryons to electroweak gauge fields radiatively influence the production and decay of the new scalar state $S_\text{B}$.  In section \ref{sec: SBProductionDecay}, we illustrate how this can make it possible to distinguish \mbox{model A} from \mbox{model VA}.


\section{Constraints on leptobaryon parameters}
\label{sec: Constraints}
Leptobaryon model A brings in a number of new parameters beyond the standard model: the gauge coupling, mass and kinetic mixing parameter
\begin{equation}
\alpha_\text{B} \equiv \frac{g_\text{B}^2}{4\pi},\,\, m_{Z_\text{B}}, \, \sin\epsilon;
\end{equation}
the baryonic Higgs mass and its mixing with the standard model Higgs
\begin{equation}
m_S,\,\theta;
\end{equation}
and the leptobaryon masses along with the Yukawa couplings to the standard model Higgs
\begin{equation}
m_\psi,\,m_\Sigma,\,m_\chi, \enspace \lambda_1,\lambda_2,\,\lambda_3,\,\lambda_4.
\end{equation}
In this section, we review some important experimental and theoretical constraints that will motivate certain simplifying assumptions on these parameters leading to simpler phenomenology in the sections to follow.


\subsection{Electroweak precision}
Because the couplings of leptobaryons to the electroweak gauge bosons in Eq.~(\ref{eq:extEWCurrentModelA}) are pure vector, they nominally make no oblique corrections at one loop ($S-S_\text{SM} = T-T_\text{SM} = 0$).  However mixing among the leptobaryons induced by the Yukawa couplings $\lambda_1 \ldots \lambda_4$ in Eq. (\ref{eq:YukawaWithH}) leads to deviations suppressed by the baryonic Higgs and leptobaryon masses \cite{Yang:2014zma}.  As mentioned near the end of Section \ref{sec:ParticleSpectrum}, we take these Yukawa couplings infinitesimally small only to permit the decay of the heavier leptobaryons to the lightest one.

Next, we consider the effect of kinetic mixing parameter $\sin(\epsilon)$ on low energy physics.  For the range of $Z_\text{B}$ masses we consider in this study ($\sim$0.5---2.5 TeV), kinetic mixing is most constrained by its effect on the relative strengths of the weak charged and neutral currents \cite{Hook:2010tw} parametrized by the $\rho$-parameter.  In the limit of small kinetic coupling,
\begin{equation}
\Delta\rho = \frac{m_W^2}{m_{Z_\text{B}}^2}\tan^2(\theta_\text{w}) \epsilon^2+ \mathcal{O}\big(\frac{m_W^4}{m_{Z_\text{B}}^4}\big) + \mathcal{O}\big(\epsilon^4\big) \,.
\end{equation}
Fits of the $\rho$-parameter to electroweak precision observables suggest a deviation from unity $\rho_0 = 1.00040 \pm 0.00024$ yielding the preferred range of
\begin{equation}\label{eq:kinMixingRange}
0.29 \lesssim \frac{\epsilon}{m_{Z_\text{B}}} \text{TeV} \lesssim 0.58\,.
\end{equation}
We shall take the kinetic mixing to zero for simplicity (consistent at the 2$\sigma$ level).

Finally, the observation of a Higgs-like resonance by the ATLAS \cite{Aad:2012tfa} and CMS \cite{Chatrchyan:2012ufa} experiments consistent with signal strengths predicted by the standard model puts a limit on any mixing that the standard model Higgs might have with other degrees of freedom.  We adopt the 95\% confidence limit on the mixing obtained by the ATLAS Collaboration \cite{ATLAS:funny}
\begin{equation}\label{eq:MixingConstraint}
|\theta| \lesssim 0.35\,.
\end{equation}


\subsection{Vacuum metastability at one loop}\label{sec:VacMetastab}
Similar to the destabilization of the standard model Higgs potential by the large top quark mass \cite{Lindner:1988ww}, the leptobaryons are also expected to destabilize the potential along the baryonic Higgs field direction, yielding a constraint on the mass spectrum.  A complete vacuum metastability analysis involving the full one-loop corrections to the scalar potential is beyond the scope of this paper.  However, we can find fairly robust constraints by studying the behavior of the potential in the leading logarithmic approximation.

We solve the one-loop $\overline{\text{MS}}$ renormalization group equation for the baryonic Higgs quartic self-coupling
\begin{multline}\label{eq:beta-function}
\mu\frac{\mathrm{d}b}{\mathrm{d}\mu} = \frac{1}{16\pi^2}\big[-4y_\psi^4-3y_\Sigma^4-y_\chi^4+2a^2+486g_\text{B}^4 \\
+2b(-54g_\text{B}^2+4y_\psi^2+3y_\Sigma^2+y_\chi^2)+20b^2\big]\,,
\end{multline}
with the initial conditions given at \mbox{$\mu_0 = (m_S + m_{Z_\text{B}})/2$}, the defining scale for the masses and couplings.  The scale $\mu=\Lambda_\text{destab.}$ at which the quartic coupling crosses zero indicates the onset of destabilization in the $S_\text{B}$ direction, and represents the scale at which new physics must enter to restore stability of the scalar potential.

We impose self-consistency on the model parameters by requiring that the potential be stable at least up to the largest mass in the theory, the leptobaryon masses $m_{\psi,\Sigma,\chi}$.  We assume degeneracy among the leptobaryons so that $y_\psi = y_\Sigma = y_\chi$, and we neglect the running of all other couplings in (\ref{eq:beta-function}) which is justified \emph{a posteriori} due to the very low scale at which $b$ crosses zero.

By retaining the dominant Yukawa coupling contributions (first three terms) in the right-hand side of (\ref{eq:beta-function}), the renormalization group equation can be readily solved in closed form yielding an approximate constraint on the particle masses
\begin{equation}\label{eq:VacStabBound}
m_{\psi,\Sigma,\chi} \, \lesssim \, 0.86 \Big(\frac{m_{Z_\text{B}}m_S}{g_\text{B}}\Big)^{1/2}\,.
\end{equation}
This approximation agrees with the numerical determination to better than 5\% and indicates that unless the gauge coupling $g_\text{B}$ is exceptionally small, fermion masses should be of the same order of magnitude as the baryonic Higgs and gauge boson  masses.  In particular, vacuum stability imposes a stringent upper limit on dark matter mass.

\subsection{On the naturalness of small parameters}\label{sec: SmallParameters}
In the current study, we take the kinetic mixing parameter $\sin(\epsilon)$ to vanish, and (for the collider study) we consider the scalar mixing angle $\theta$ reaching $\sim10^{-5}$ in magnitude.  The smallness of these parameters is not associated with the emergence of any symmetry, and in this section we ask whether these choices are consistent with the radiative corrections.  

The kinetic mixing angle $\sin(\epsilon)$ is the coefficient of a marginal operator $B_{\mu\nu}Z_\text{B}^{\mu\nu}$, and is therefore logarithmically sensitive to the UV scale.  Without information about UV physics, no meaningful statements can be made about the size of radiative corrections.  However, if (setting aside the issue of vacuum stability) grand unification as suggested in \cite{Perez:2014kfa} is assumed and the U(1) hypercharge and baryon factors are orthogonal, $\sin(\epsilon)$ can be estimated based on the renormalization group evolution equation starting from $\sin(\epsilon)=0$ at the unification scale\footnote{See Table I of \cite{Perez:2014kfa}} $\Lambda_\text{GUT} = 1.24\times10^{12}$ GeV.
The quarks and charged leptobaryons ($\psi$ and $\Sigma$) contribute to the beta function at one-loop order, given by
\begin{equation}
\mu \frac{\mathrm{d}(\sin\epsilon)}{\mathrm{d}\mu} =  \frac{1}{16\pi^2}\textstyle \big[-\frac{8}{3}g' g_\text{B}+(\frac{15}{2} g'^2 + \frac{53}{3}g_\text{B}^2)\sin\epsilon\big]\,.
\end{equation}
By evolving the kinetic mixing parameter down to the leptobaryon masses ($\sim$TeV scale), we find $\sin(\epsilon) \approx 0.06$.  We emphasize that this estimate should be taken with a grain of salt given that vacuum instability sets in at a much lower scale.  New physics that stabilizes the potential is likely to modify the beta function.

The scalar sector mixing angle $\theta$ is derived from the Higgs portal coupling $a$, which is also the coefficient of a marginal operator $H^\dag H S_\text{B}^* S_\text{B}$.  In principle, a renormalization group evolution from the unification scale would provide an estimate at the TeV scale.  But in this instance, more information about the structure of unification in the scalar sector is needed to make such an estimate.  However, we mention that within the low energy leptobaryon model, renormalization effects are proportional to itself at the one-loop level.  Independent renormalization starts at the two-loop level, given by
\begin{equation*}
-i\Sigma_{HS} = \parbox{3cm}{\includegraphics[width=3cm]{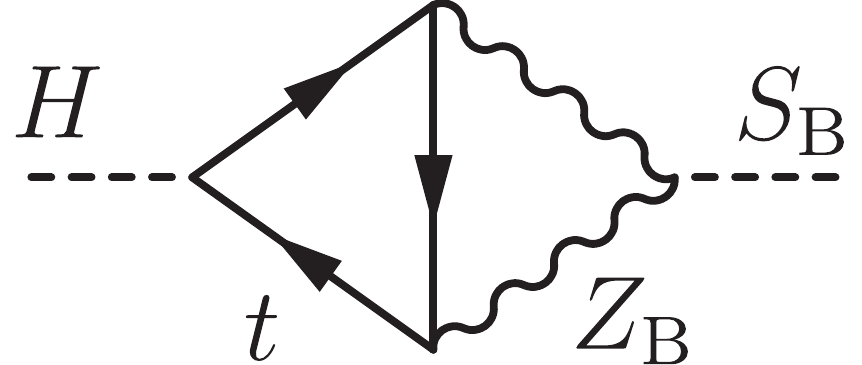}}\,.
\end{equation*}
Therefore, we expect that the mixing angle is fairly stable against quantum corrections. 



\section{Baryogenesis}\label{sec:Baryogenesis}
In \cite{Perez:2014qfa,Perez:2013tea}, it was shown that if $\text{U(1)}_\text{L}$ is broken at a much higher scale, the decay of right-handed Majorana neutrinos can create an initial $B-L$ asymmetry in the early Universe.   The leptobaryons $\psi$ and $\Sigma$ carry weak isospin, and therefore modify the 't Hooft operator \cite{Perez:2014qfa} associated with the sphaleron
\begin{align}
(Q Q Q L)^3 \bar{\Psi} \Psi \Sigma^4\,.
\end{align}
Above electroweak and baryon symmetry breaking scales, the modified sphalerons process the initial asymmetry $\Delta(B-L)_\text{SM}$ into a final baryon asymmetry $\Delta B_f^\text{SM}$, with a conversion factor given by
\begin{align}\label{eq:lbEqBaryonAymmetric}
\Delta B^\text{SM}_f = \frac{32}{99} \Delta(B-L)_\text{SM} \approx 0.32 \ \Delta(B-L)_\text{SM} \,.
\end{align}
This is slightly different from the standard model conversion factor first computed in \cite{Harvey:1990qw}
\begin{align}\label{eq:smEqBaryonAymmetric}
\Delta B^\text{SM}_f = \frac{28}{79} \Delta(B-L)_\text{SM} \approx 0.35 \ \Delta(B-L)_\text{SM} \,.
\end{align}
Provided that $H$ acquires a {\sc vev} earlier than (or at the same time as) $S_\text{B}$ does in the history of the Universe, the sphaleron processes decouple, and the relationship in (\ref{eq:lbEqBaryonAymmetric}) is frozen in.  In view of the hierarchy of scales assumed in this model, it seems more plausible that $H$ and $S_\text{B}$ acquire {\sc vev}s simultaneously.

We mention here an important caveat\footnote{We thank David Morrissey for alerting us to this possibility.} that was not fully appreciated in \cite{Perez:2014qfa}.  If $S_\text{B}$ acquires a {\sc vev} \emph{before} $H$ does, the Dirac mass term for $\Psi$ imposes the chemical equilibrium condition $\mu_{\Psi} = \mu_{\bar\Psi}$.  Then, the conversion factor for the final baryon asymmetry reverts back to the standard model one in (\ref{eq:smEqBaryonAymmetric}).


\section{Leptobaryonic cold dark matter}
\label{sec:DarkMatter}
We begin our phenomenological exploration of leptobaryon model A by investigating the properties of the dark matter candidate.  There are three electrically neutral leptobaryons, $\psi^0, \Sigma^0$ and $\chi$, each one of which could in principle account for the observed dark matter abundance. 
In this paper, we take the isosinglet $\chi$ as the lightest stable leptobaryon to serve as the dark matter candidate, and defer the exploration of the isotopically charged leptobaryons $\psi^0$ and $\Sigma^0$ as dark matter to future study.

Note that since $\chi$ is a Majorana fermion, its coupling to $Z_\text{B}$ is purely axial [see (\ref{eq:extBaryonCurrentModelA})], and is in contrast with the dark matter candidate of model VA which also couples vectorially [see (\ref{eq:extBaryonCurrentModelVA})].  This leads to major qualitative differences in the dark matter phenomenology with respect to model VA, largely because several possible scattering channels are velocity suppressed.

\subsection{Dark matter relic abundance and direct detection}
\label{sec: DM Annihilation}
We assume that the dark matter is produced by standard thermal freeze-out.  The dark matter abundance in the Universe is governed by the Lee-Weinberg equation \cite{Lee:1977ua}
\begin{align}
\frac{\mathrm{d}Y}{\mathrm{d}x}=Z(x)[Y^2_{\text{eq}}(x)-Y^2(x)]\,,
\label{eq:Lee-Weinberg}
\end{align}
where $Y(x)$ is the dark matter number density normalized by the entropy density, and $x=m_\chi/T$ is the scaled plasma temperature.  The factor $Z(x)$ is given by
\begin{align}
Z(x)= \sqrt{\frac{\pi}{45}}\frac{m_\chi m_\text{P}}{x^2}\sqrt{g_*} \langle v \sigma \rangle\,,
\end{align}
where the thermally averaged cross section is \cite{Gondolo:1990dk}
\begin{align}
\langle v \sigma\rangle = \int_{4m_\chi}^\infty \mathrm{d}s \frac{\sqrt{s} (s-4m^2_\chi)\sigma(s)K_1(\frac{x \sqrt{s}}{m_\chi})x}{8 m^5_\chi K^2_2(x)} \,.
\end{align}
The large $x$ limit of the solution to (\ref{eq:Lee-Weinberg}) determines the thermal relic abundance.

We numerically integrate the thermally averaged cross section, retaining all orders in the velocity expansion to accurately capture the resonant and threshold behaviors, and we solve (\ref{eq:Lee-Weinberg}) in the freeze-out approximation \cite{Bender:2012gc} to give
\begin{align}\label{eq:dmapprox}
Y(\infty)=\textstyle\big[\frac{1}{Y(x_f)}+{\int_{x_f}^{\infty} Z(x)\mathrm{d}x} \big]^{-1}\,,
\end{align}
with the freeze-out temperature $x_f$ determined by matching.  
To improve computational speed, we neglect the first term in (\ref{eq:dmapprox})  which leads to an overestimation of the relic abundance by no more than 5\%.  The thermal relic abundance is then obtained from
\begin{align}
\Omega_\text{DM} = \frac{m_\chi s_0}{\rho_c}Y(\infty) \,,
\end{align}
where $s_0 = 2970/\text{cm}^3$ is the entropy density today and $\rho_c = 1.05394\times 10^{-5} h^2\, \text{GeV}/\text{cm}^3$ is the critical density.

In Table \ref{tab: FeynmanDiagrams}, we list the tree-level dark matter annihilation channels, all of which are taken to compute the dark matter relic density.  Channels going directly to standard model particles such as $\chi \chi \to H S_\text{B}$ and $\chi \chi \to H H$ are omitted because they are suppressed by at least 1 order of magnitude owing to the collider constraint on the scalar mixing angle $|\theta| < 0.35$.  To facilitate the identification of velocity suppressed amplitudes, the lowest contributing partial waves are also given.  Amplitudes for channels with an initial $P$ (or higher) wave will be down by at least one power of dark matter velocity.  

\begin{table}[t]
\begin{tabular}{cc}
	\multicolumn{2}{c}{Resonant annihilation channels}\\
	\hline \multicolumn{1}{l}{$\chi\chi\rightarrow \bar{q}q$}\\[1mm]
	\includegraphics[width=15mm]{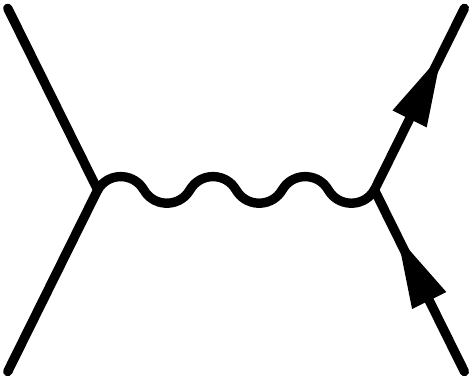} \\
		$^3P_1^+ \rightarrow 1^- \rightarrow \bar{q}q$\\[3mm]

	\multicolumn{2}{c}{Non-resonant annihilation channels}\\[1mm]
	 \hline\multicolumn{1}{l}{$\chi\chi\rightarrow Z_\text{B}S_\text{B}$}\\[1mm]
	\includegraphics[width=13.6mm]{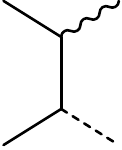} \enspace\enspace   
	\includegraphics[width=13.6mm]{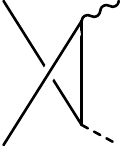}
  & \includegraphics[width=15mm]{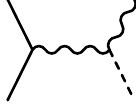}\\
  $^1S^-_0 \rightarrow {}^3P_0^+$ & $^1S_0^- \rightarrow 0^+ \rightarrow {}^3P^+_0$\\[2mm]
  
  \multicolumn{1}{l}{$\chi\chi\rightarrow Z_\text{B}Z_\text{B}$}\\[1mm]
	\includegraphics[width=13.6mm]{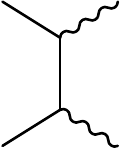} \enspace\enspace  
	\includegraphics[width=13.6mm]{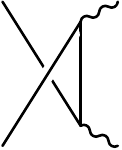}
  & \includegraphics[width=15mm]{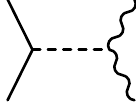} \\
  $^1S^-_0 \rightarrow {}^3P_0^-$ & $^3P_0^+ \rightarrow 0^+ \rightarrow \Big\{\!\!\begin{array}{c}{}^1S^+_0 \\ {}^5D_0^+\end{array}\!\!\Big\}$\\[2mm]
  
  \multicolumn{1}{l}{$\chi\chi\rightarrow S_\text{B}S_\text{B}$}\\[1mm]
  \includegraphics[width=13.6mm]{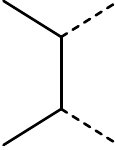} \enspace\enspace   
	\includegraphics[width=13.6mm]{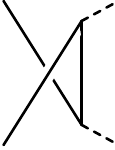}
  & \includegraphics[width=15mm]{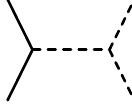} \\
  
  $^3P^+_0 \rightarrow {}^1S_0^+$ & $^3P_0^+ \rightarrow 0^+ \rightarrow {}^1S^+_0$\\[1mm]
\hline
\end{tabular}
\caption{Set of dark matter annihilation channels in the limit of vanishing scalar $H$-$S_\text{B}$ mixing angle $\theta$, with the lowest nonvanishing partial waves indicated.  Channels with an entering $S$-wave ($\chi\chi\rightarrow Z_\text{B}S_\text{B}$ and $\chi\chi\rightarrow Z_\text{B}Z_\text{B}$) are not velocity suppressed.}
\label{tab: FeynmanDiagrams}
\end{table}

\begin{figure}[!htb]
	\centering
		\includegraphics[width=0.45\textwidth]{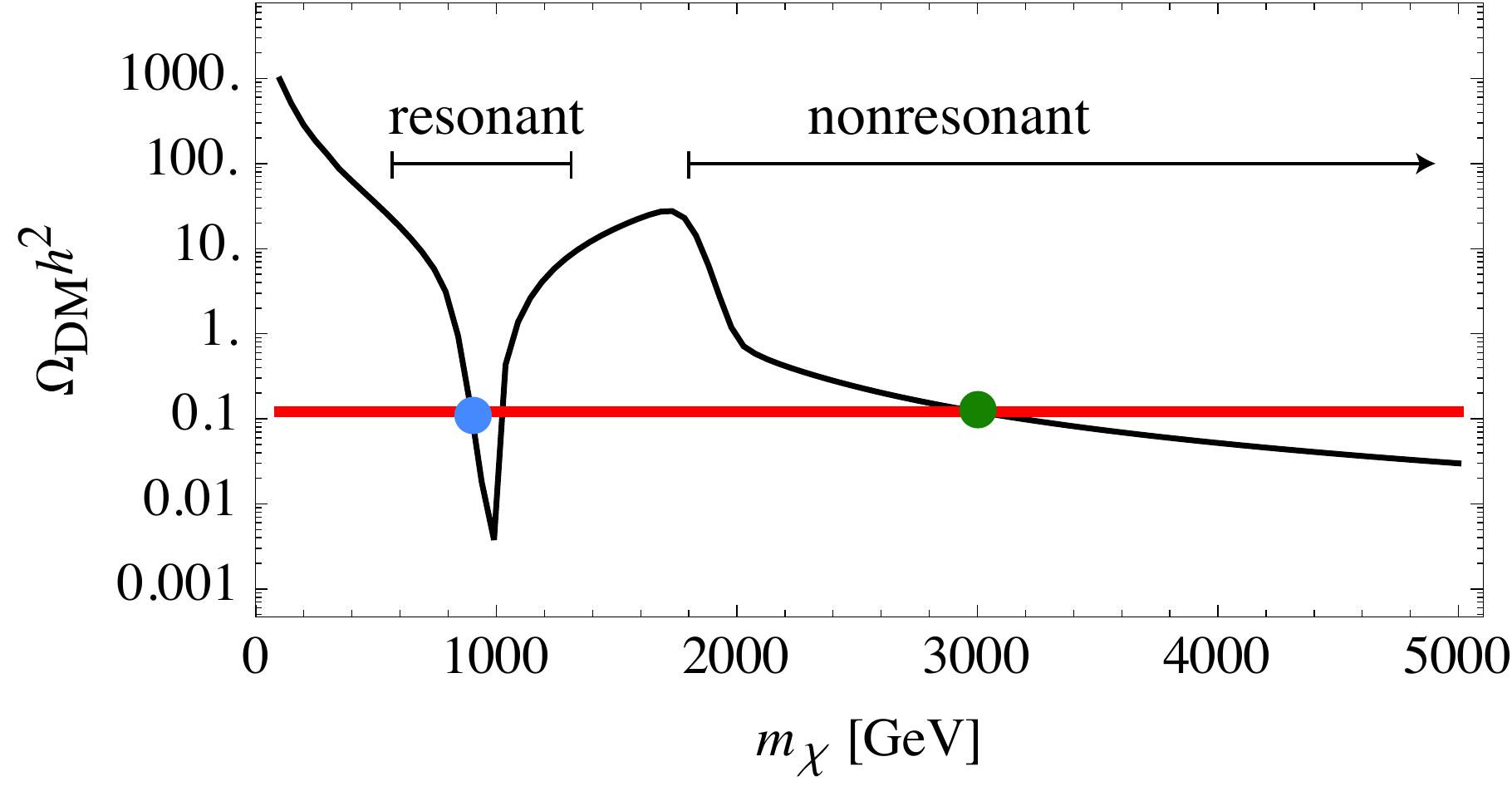}
	\caption{Dark matter relic abundance (black) as a function of the dark matter mass $m_\chi$ for fixed $\alpha_\text{B}$, $m_{Z_\text{B}}$, $m_S$ and $\theta$. The red line corresponds to the measured abundance $\Omega_\text{DM} h^2 = 0.12$.  The regions labeled `resonant' and `nonresonant' indicate the principal dark matter annihilation channels (see Table \ref{tab: FeynmanDiagrams}) that determine the relic abundance.}
	\label{fig: DarkMatterRelicAbundance}
\end{figure}

In Fig. \ref{fig: DarkMatterRelicAbundance}, we plot the dark matter relic abundance $\Omega_\text{DM}h^2$ as a function of dark matter mass $m_\chi$ for a typical choice of model parameters.  The observed abundance at $\Omega_\text{DM} h^2 = 0.12$ \cite{Ade:2013zuv} is indicated by the red line, and intersections determine values of $m_\chi$ that saturate the relic density.  The two intersections to the very left correspond to comparatively light dark matter ($m_\chi \approx m_{Z_\text{B}}/2$), and are determined by the resonant annihilation channel $\chi\chi \rightarrow Z_\text{B} \rightarrow q\bar{q}$.  The single intersection to the right gives the correct relic abundance for heavier dark matter masses, and arises from nonresonant annihilation which is usually dominated by the velocity unsuppressed channels $\chi \chi \rightarrow Z_\text{B} S_\text{B}$ and $\chi\chi \rightarrow  Z_\text{B} Z_\text{B}$.  To simplify the phenomenology, we will later separate the discussion corresponding to the two distinct mass ranges of the dark matter candidate.\\

\begin{figure}
	\centering
		\begin{tabular}{cc}
	\multicolumn{2}{l}{\includegraphics[width=56.0mm]{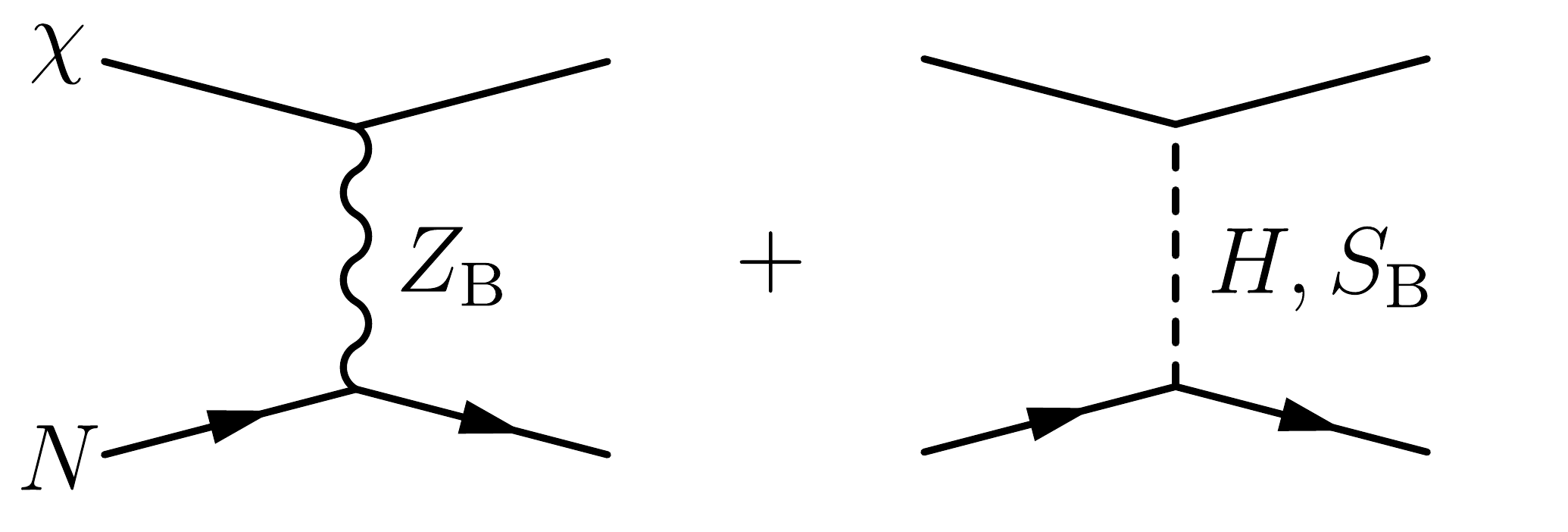}}\\
  \hspace{5mm}$^{1}S^-_0 \rightarrow {}^{3}P_0^+$ & \hspace{5mm} $^{1,3}S_0^- \rightarrow {}^{1,3}S_0^-$
\end{tabular}
	\caption{Processes contributing to the spin-independent direct detection cross section, with the lowest nonvanishing partial waves indicated. Scattering mediated by $Z_\text{B}$ exchange is velocity suppressed because it is a (parity-change) $\Delta \ell = 1$ transition, while scattering mediated by $H$ and $S_\text{B}$ exchanges is mixing angle suppressed.}
	\label{fig: DirectDetection}
\end{figure}

At tree-level, the dark matter--nucleon scattering channels for direct detection are the $t$-channel $Z_\text{B}$, $H$ and $S_\text{B}$ exchanges.  The Feynman graphs are shown in Fig. \ref{fig: DirectDetection} together with their lowest contributing partial waves.  The $Z_\text{B}$ exchange is velocity suppressed since it involves a change in the orbital angular momentum on account of the axial coupling to dark matter.  On the other hand, the $H$ and $S_\text{B}$ exchange is suppressed by the mixing angle.

At leading order in the velocity expansion, the lab frame spin-independent direct detection cross sections, in the limit $m_\chi \gg m_N$, are
\begin{align}
\sigma_\text{SI}(Z_\text{B})=&18\pi \alpha_\text{B}^2 \frac{3m_N^2}{m_{Z_\text{B}}^4} v^2\,,
\label{eq: DirectDetectionZB}\\
\nonumber \sigma_\text{SI}(S_\text{B}) =& \frac{72 G_F \alpha_\text{B}}{\sqrt{2}}\sin^2\theta \cos^2\theta \\ 
&\hspace{1cm}\times\frac{m_\chi^2 m_N^4}{m_{Z_\text{B}}^2} \Big(\frac{1}{m_H^2}-\frac{1}{m_S^2}\Big)^2 F_N^2, 
\label{eq: DirectDetectionSB}
\end{align}
where $m_N$ is the nucleon mass and $F_N = 0.32$ is the effective nucleon matrix element taken from lattice calculations \cite{Hisano:2010ct}.  The total direct detection cross section is given by the sum $\sigma_\text{SI} = \sigma_\text{SI}(Z_B) + \sigma_\text{SI}(S_B)$.  Orthogonality of the partial waves forbids an interference term.

Altogether, the dark matter phenomenology principally depends on five free parameters
\begin{align}\label{eq:DMmodelparameters}
m_{Z_\text{B}},\, m_S,\, m_\chi,\, \alpha_\text{B},\, \theta \,.
\end{align}

\subsection{Resonant dark matter}
\label{Resonant Annihilation}

\begin{figure}
	\centering
		\includegraphics[width=\columnwidth]{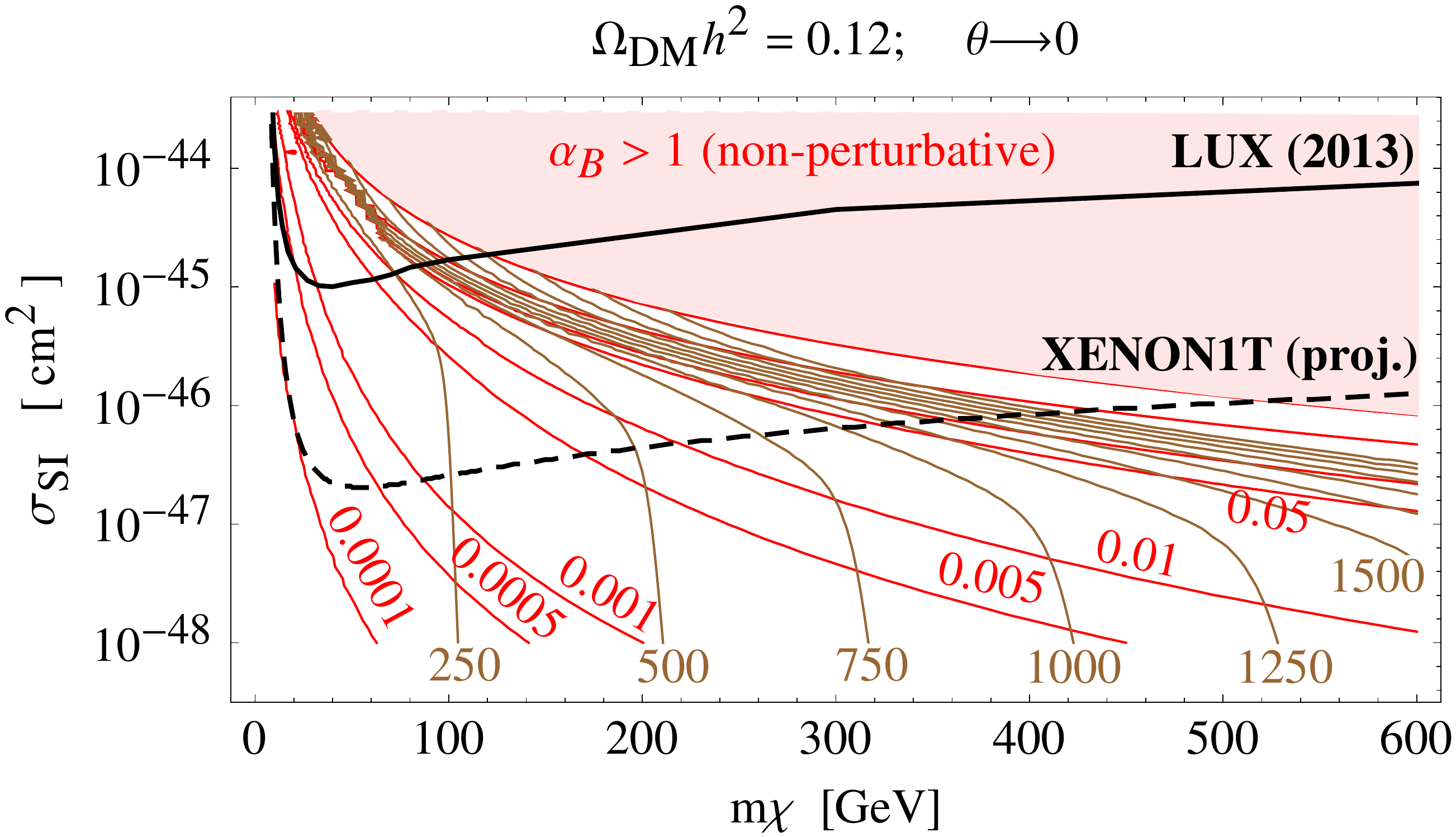}
	\caption{Contours of fixed $m_{Z_\text{B}}$ (brown lines, in GeV) and $\alpha_\text{B}$ (red lines) in the $m_\chi$--$\sigma_\text{SI}$ plane to achieve dark matter relic abundance saturation, for $m_\chi$ in the resonant annihilation region of Fig. \ref{fig: DarkMatterRelicAbundance} (blue dot).  In the upper-right region, a nonperturbatively large coupling is needed.  The limit $\theta\rightarrow 0$ is taken to decouple the scalar exchanges from direct detection.  The current LUX (2013) bound \cite{Akerib:2013tjd} and the XENON1T projections \cite{Aprile:2012zx} are shown.}
	\label{fig: Resonant_DM}
\end{figure}

Motivated by the vacuum stability bound (\ref{eq:VacStabBound}), we start by assessing the prospects for the relic abundance being saturated by light dark matter due to the resonant (velocity suppressed) annihilation channel\footnote{The resonant annihilation channel $\chi\chi \rightarrow H/S_\text{B} \rightarrow q \bar{q}$ is both velocity \emph{and} mixing angle suppressed.} $\chi \chi \to Z_B \to q \bar{q}$ in the early Universe.  The annihilation cross section is given by
\begin{equation}
\sigma = 9 \pi \alpha_\text{B}\sqrt{1-\frac{4m_\chi^2}{s}}\,\frac{m_{Z_\text{B}}\Gamma_{Z_\text{B}}\text{Br}(Z_\text{B}\rightarrow \sum \bar{q}q)}{(s-m^2_{Z_\text{B}})^2 + m^2_{Z_\text{B}}\Gamma^2_{Z_\text{B}}}\,,
\end{equation}
where the $Z_\text{B}$ total width $\Gamma_{Z_\text{B}}$ is evaluated at $s = m_{Z_\text{B}}^2$ given in (\ref{eq:zbwidths}).
The annihilation cross section depends only on $\alpha_\text{B}$, $m_{Z_\text{B}}$ and $m_\chi$.  For the moment, we decouple the scalar particles $H$ and $S_\text{B}$ from the problem by taking $\theta\rightarrow 0$ so that the only direct detection channel is via the $t$-channel $Z_\text{B}$ exchange\footnote{Direct detection experiments report a limit on $\sigma_\text{SI}$ under the assumption that to leading order it is velocity-independent.  Since the $Z_\text{B}$ exchange gives a cross section that is velocity-dependent to leading order, we scale the $\sigma_\text{SI}(Z_\text{B})$ in (\ref{eq: DirectDetectionZB}) by the isothermal halo model of dark matter distribution \cite{Freese:1987wu} to obtain an effective cross section $\sigma_0 \sim \sigma_1^\text{SI} \langle v\rangle/\langle v^{-1}\rangle$.  The net effect is to use an effective lab-frame velocity of $v\approx0.0093c$.}.  We return to discuss the effects of restoring the scalar exchanges below.

In this limit dependence on $m_S$ and $\theta$ in the $\chi$-$N$ cross section drops out, and a positive signal by direct detection experiments can fully determine the three remaining parameters in (\ref{eq:DMmodelparameters}).  For a given dark matter mass $m_\chi$, the direct detection cross section in (\ref{eq: DirectDetectionZB}) is inverted to fix $\alpha_\text{B}$ as a function of $m_{Z_\text{B}}$.  The gauge boson mass $m_{Z_\text{B}}$ is then determined by requiring saturation of the observed relic density.  For illustrative purposes, we take the lower dark matter mass as shown in Fig. \ref{fig: DarkMatterRelicAbundance} by a blue dot.

In Fig. \ref{fig: Resonant_DM}, we display the contours of constant coupling strength $\alpha_\text{B}$ (in red) and gauge boson mass $m_{Z_\text{B}}$ (in brown) in the $m_\chi$--$\sigma_\text{SI}$ plane assuming relic abundance saturation.  
The shape of the brown $m_{Z_\text{B}}$ contours reflects the shape of the relic abundance curve in Fig. \ref{fig: DarkMatterRelicAbundance} in the resonant region.  The upper-right area (large $m_\chi$ and $\sigma_\text{SI}$) is where a nonperturbatively large $\alpha_\text{B}$ is needed to achieve relic abundance saturation.  The strongest direct detection cross section bound by LUX (2013) \cite{Akerib:2013tjd} currently excludes comparatively light dark matter $m\chi \lesssim 150$ GeV assuming perturbative couplings and $\theta=0$.  Projections by the XENON1T Collaboration \cite{Aprile:2012zx} indicate sensitivity to larger dark matter masses reaching $m_\chi \sim 500$ GeV.  We find that for nonvanishing $\theta$, the velocity unsuppressed $H$ and $S_\text{B}$ exchanges (see Fig. \ref{fig: DirectDetection}) contribute to direct detection, but do not lead to significantly stronger exclusions.



\subsection{Nonresonant dark matter}
\label{Non-Resonant Annihilation}

We now turn to the case of heavy dark matter (corresponding to the nonresonant region in Fig. \ref{fig: DarkMatterRelicAbundance}) where $m_\chi \gtrsim m_{Z_\text{B}},m_S$. We find that a much larger region in parameter space is allowed, and that current experimental bounds on the spin-independent direct detection cross section provide limits on the mixing angle $\theta$ that are more stringent than from the Higgs signal strength measurements at the LHC, which we discuss here.

The annihilation channels involving the baryonic Higgs are open in this scenario, and play a crucial role in determining the relic abundance.  Thus, the dark matter phenomenology is influenced by all five parameters in (\ref{eq:DMmodelparameters}).  At leading order in the velocity expansion, the thermally averaged nonresonant annihilation cross sections are
\begin{align}
\label{eq:thermalCrossSectionZBSB}
\langle v\sigma(\chi \chi \rightarrow Z_\text{B} S_\text{B}) \rangle &= \frac{81\alpha_\text{B}^2 \pi \text{cos}^2\theta}{64\, m_\chi^4 m_{Z_\text{B}}^4}\lambda(4m_\chi^2,m_{Z_\text{B}}^2,m_{S}^2)^{3/2}\\
\label{eq:thermalCrossSectionZBZB}
\langle v\sigma(\chi\chi \rightarrow  Z_\text{B} Z_\text{B}) \rangle &= \frac{81 \pi \alpha^2_\text{B}}{4 m_\chi }\frac{(m_\chi^2-m_{Z_\text{B}}^2)^{3/2}}{(m_{Z_\text{B}}^2-2m_\chi^2)^2} \,,
\end{align}
where $\lambda(a,b,c)=a^2+b^2+c^2-2ab-2ac-2bc$.

\begin{figure*}
\includegraphics[width=0.9\columnwidth]{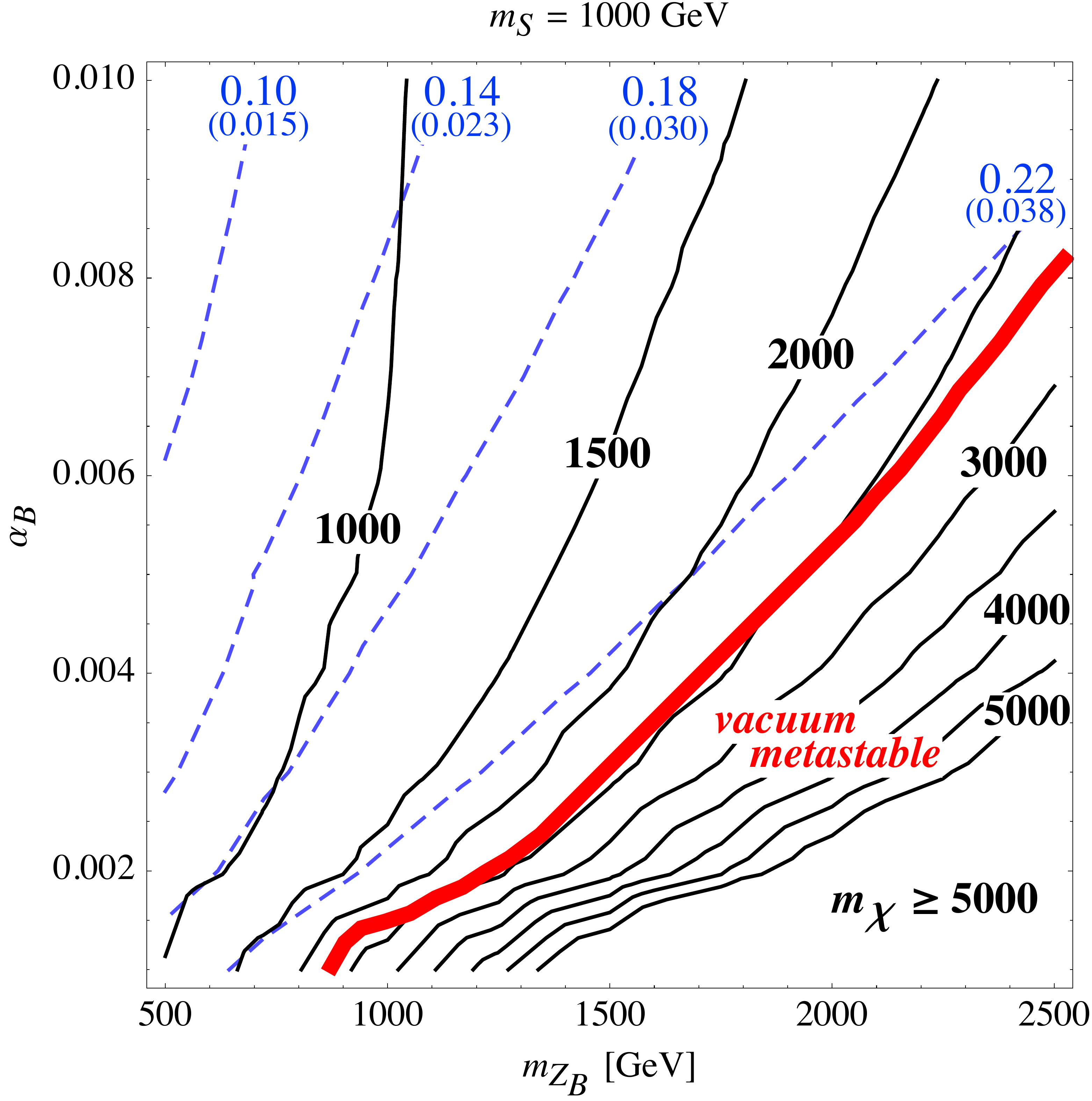}\qquad
\includegraphics[width=0.9\columnwidth]{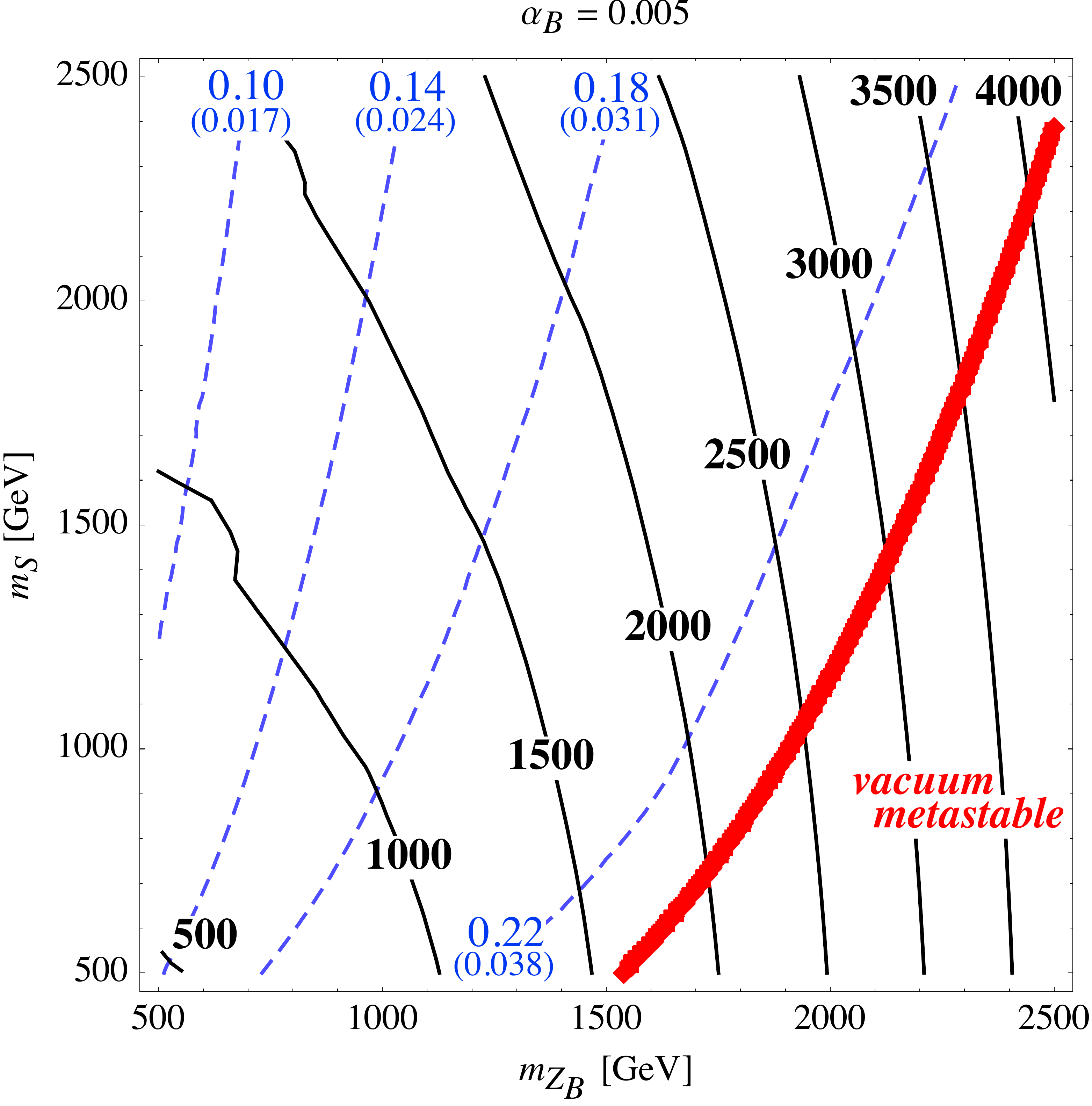}
\caption{Contours of constant dark matter mass $m_\chi$ (black, in GeV) saturating the thermal relic abundance in the `nonresonant' annihilation region of Fig. \ref{fig: DarkMatterRelicAbundance} (green dot) with the approximation $\theta=0$.  Dashed blue contours indicate upper limits on the Higgs mixing angle $\theta$ obtained from direct detection spin-independent cross section bounds.  Values outside parenthesis indicate current limits based on the LUX (2013) data \cite{Akerib:2013tjd}, and the values within parenthesis are the forecasts based on XENON1T projections \cite{Aprile:2012zx}.  The red line demarcates the region below which the vacuum is radiatively unstable at a scale lower than the dark matter mass (see Section \ref{sec:VacMetastab}). Left panel $m_{S} = 1000$ GeV fixed, and right panel $\alpha_\text{B} = 0.005$ fixed.}\label{fig: DMPlots}
\end{figure*}

In view of the LHC upper bound on $\theta$, we begin by taking $\theta=0$ in the annihilation formulas (this approximation will later be shown to be very good).  Then, for fixed values of $m_{Z_\text{B}}$, $m_S$ and $\alpha_\text{B}$, we determine the dark matter mass $m_\chi$ by requiring saturation of the relic abundance.  These values are shown as solid black contours in Fig. \ref{fig: DMPlots} with $m_S = 1$ TeV (left) and $\alpha = 0.005$ (right).

\begin{figure}
	\centering
		\includegraphics[width=0.7\columnwidth]{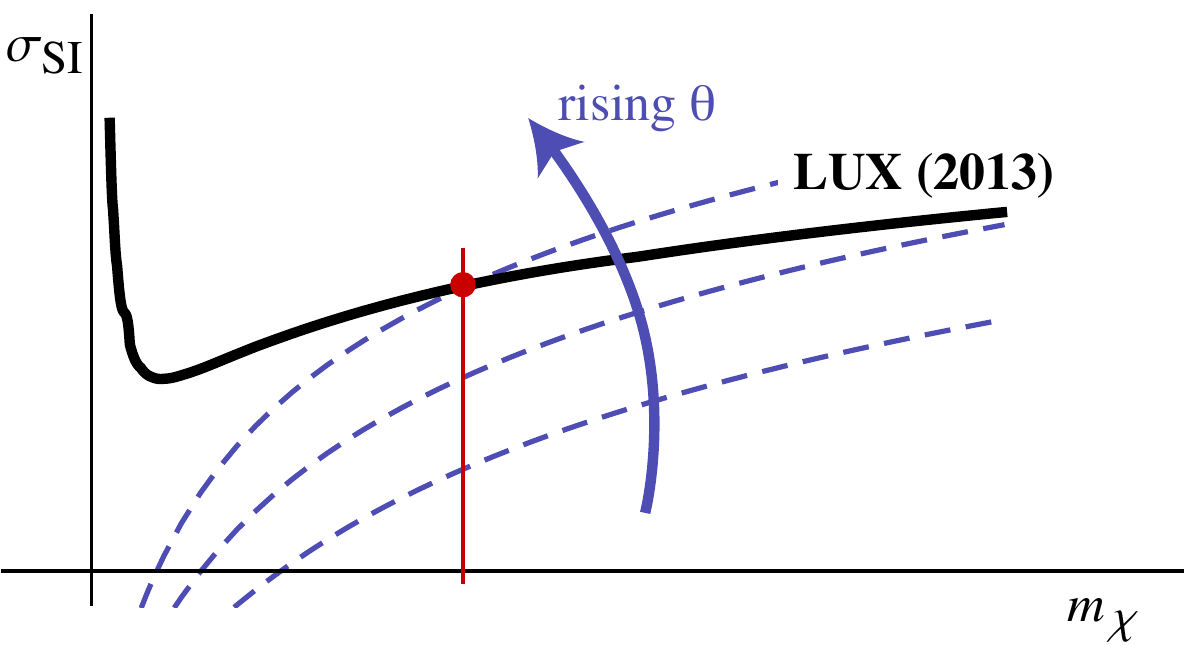}
	\caption{The spin-independent direct detection cross section $\sigma_\text{SI}$ (dashed blue) as a function of the dark matter mass $m_\chi$ for several values of the Higgs mixing angle $\theta$, for fixed $m_{Z_\text{B}}$, $m_S$ and $\alpha_\text{B}$.  The LUX (2013) bound is shown in black.  At the dark matter mass $m_\chi$ determined by relic density saturation (red line), the upper limit on $\theta$ is obtained when $\sigma_\text{DD}$ intersects the LUX bound.}
	\label{fig: DirectDetectionMix}
\end{figure}

The direct detection scattering cross section is now dominated by the $t$-channel $H$ and $S_\text{B}$ exchanges given by Eq. (\ref{eq: DirectDetectionSB}), with the velocity suppressed $Z_\text{B}$ contribution being negligibly small.  The dependence of the cross section on the scalar mixing angle is particularly important.  In Fig. \ref{fig: DirectDetectionMix}, the behavior of the spin-independent cross section (in dashed blue) is sketched for several values of $\theta$ along with the LUX (2013) \cite{Akerib:2013tjd} upper limit in black.  For the value of $m_\chi$ determined by relic abundance saturation (red vertical line), direct detection null results are used to find an upper bound on the allowed mixing angle, determined by the intersection (indicated by a red dot). 

The contours of constant upper limits on the Higgs mixing angle from the LUX (2013) data are shown as dashed blue lines in Fig. \ref{fig: DMPlots} with labels outside parenthesis.  The weakest upper limit is around $\theta = 0.22$, and justifies \emph{a posteriori} the approximation $\theta \approx 0$ originally taken in the annihilation formulas.  This procedure can be viewed as the first step of a rapidly convergent iterative process to accurately determine the upper bound on $\theta$ set by direct detection experiments.
Shown in parenthesis on the same contours are the expected limits on the mixing angle derived from the XENON1T projections, and are significantly stronger than the LHC bound ($|\theta| \lesssim 0.35$).


The rise in dark matter mass $m_\chi$ required for saturation as the coupling $\alpha_\text{B}$ is lowered (left panel of Fig. \ref{fig: DMPlots}) can be qualitatively understood from Fig. \ref{fig: DarkMatterRelicAbundance}.  A decrease in $\alpha_\text{B}$ implies smaller annihilation cross section, which leads to larger relic density corresponding to an overall upward shift in the black curve in Fig. \ref{fig: DarkMatterRelicAbundance}.  This then displaces the intersection point (green dot) to the right implying heavier dark matter mass.   The rise in $m_\chi$ with increasing $m_{Z_\text{B}}$ and $m_{S}$ (right panel of Fig. \ref{fig: DMPlots}) is more readily understood as requiring the nonresonant annihilation channels to remain open.

With such heavy dark matter $\chi$ and even heavier charged leptobaryons $\psi$ and $\Sigma$, the bounds derived from one-loop vacuum stability in Section \ref{sec:VacMetastab} become important.  Displayed in the same plots is the boundary (red line) where the vacuum becomes unstable at a scale equal to the dark matter mass.  In the region to the right of the red line, the vacuum becomes unstable at a scale below the dark matter mass, and signals the presence of new physics to restore vacuum stability.

Despite bounds from one-loop vacuum metastability, the allowed region of parameter space for larger dark matter masses $m_\chi \gtrsim m_{Z_\text{B}} \simeq m_S$, is much wider than that for lighter dark matter masses $m_\chi \lesssim m_{Z_\text{B}}/2$ .  In addition, direct detection experiments are becoming more sensitive to the Higgs mixing angle and we expect that future experiments will be able to tighten the limit on $\theta$ significantly more strongly than that from the Higgs signal strength measurements at the LHC.

\subsection{Upper bound on $\text{U(1)}_\text{B}$ breaking scale?}
\label{Bound}
We close this section by revisiting the arguments in \citep{Duerr:2014wra}, in the context of leptobaryon model VA, leading to an absolute upper bound on the baryon symmetry breaking scale ($\sim 35$ TeV) that is much lower than the one derived from unitarity ($\sim 300$ TeV) \cite{Griest:1989wd}.  Such a bound would be welcome, as it would imply falsifiability in the near future.  In an attempt to derive a similar bound in the present model, we found that the arguments are based on an invalid implicit assumption, and upon further scrutiny the upper limit is lost.

The implicit assumption made in the argument is that among the possible dark matter annihilation channels, the resonant channel $\chi\bar\chi \rightarrow Z_\text{B} \rightarrow q\bar{q}$ is always the most efficient.  It is then shown that as the symmetry breaking scale, \emph{i.e.} the $Z_\text{B}$ mass, is parametrically increased, the thermal annihilation cross section for this channel drops.  Eventually, the magnitude of the cross section becomes too low, unavoidably leading to an overclosed Universe.  The value of $m_{Z_\text{B}}$ when this occurs evidently becomes the upper limit on the baryon symmetry breaking scale.

However, we point out that even for large $m_{Z_\text{B}}$, the nonresonant annihilation cross sections can be made arbitrarily large by increasing $\alpha_\text{B}$ thereby guaranteeing sufficient dark matter annihilation.  One may worry that a nonperturbatively large $\alpha_\text{B}$ or $y_\chi$ is needed, or that unitarity must be violated before reaching sufficient annihilation.  However, this concern has already been addressed in \cite{Griest:1989wd}.

The arguments above also rely on $\chi$ being the lightest leptobaryon.  It is possible for the isodoublet $\psi$ to be the lightest leptobaryon, with the electrically neutral component $\psi^0$ as the dark matter candidate.  Because of its electroweak charges, many additional annihilation channels are open, and it is not difficult to imagine a scenario with high scale $\text{U(1)}_\text{B}$ breaking and where $\psi^0$ forms a subdominant part of multicomponent dark matter.

We mention that, within the minimal leptobaryon models, it may be possible to use bounds from vacuum metastability in conjunction with an analysis of the annihilation channels to place an upper bound on the $\text{U(1)}_\text{B}$ breaking scale.  However, given that new physics is expected to enter, the bound cannot be viewed as being absolute.  Nevertheless, we believe it can provide a useful range of scales for experimentalists which makes it worthwhile for further investigation.

\section{Minimal leptobaryons at the LHC}
\label{sec: Colliders}
To definitively test this model, one would ideally like to produce the new leptobaryons $\psi$, $\Sigma$, $\chi$ at a collider and measure their electroweak quantum numbers.  However, the requirement imposed by dark matter relic abundance discussed in the previous section suggests that the lightest leptobaryon $\chi$ is more likely to have a mass in the TeV range.  On account of the accidental global $Z_2$ symmetry, leptobaryons must be produced in pairs.  This leads to phase space suppression in the production rates, and makes their direct observation more difficult.

However, it is possible to reveal the underlying baryonic symmetry breaking mechanism by producing and studying the gauge boson $Z_\text{B}$ and the Higgs boson $S_\text{B}$, which will be discussed in this section.  Although these new bosons are not specific to this model, for very small values of the Higgs mixing angle $|\theta|\lesssim 10^{-3}$, we find that it may be possible to indirectly observe the charged leptobaryons based on their imprints left on the loop-mediated decay modes of $S_\text{B}$.

\subsection{Production of $Z_\text{B}$ and limits on $\alpha_\text{B}$}
\begin{figure}
\includegraphics[width=\columnwidth]{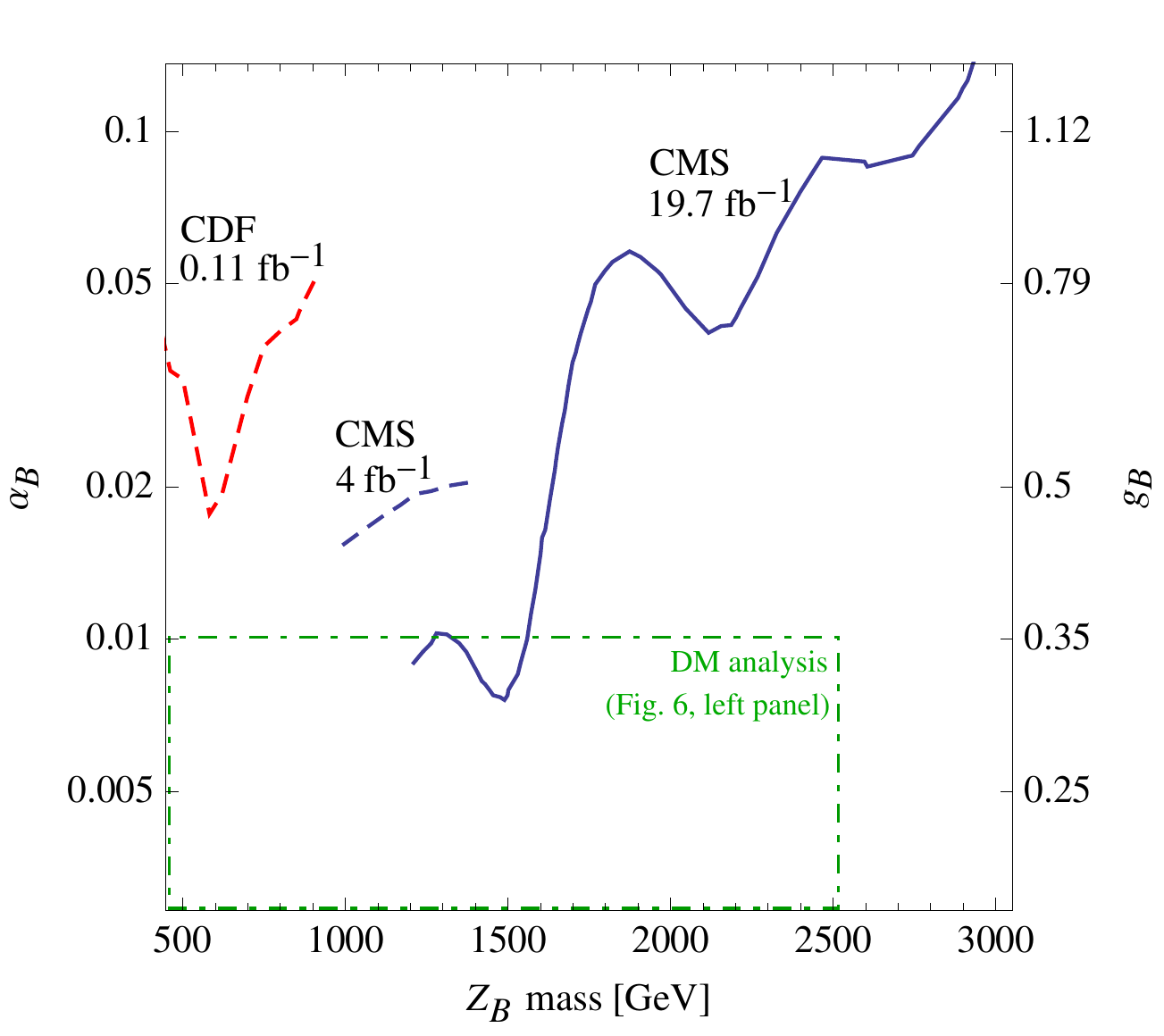}
\caption{Upper limits on $\alpha_\text{B}$ as a function of $Z_\text{B}$ mass drawn from the absence of a dijet resonance signal.  Note that $\alpha_\text{B}$ is on a logarithmic scale.}\label{fig:CMSDijetExclusion}
\end{figure}
We identify two dominant production modes of $Z_\text{B}$ at the LHC: the $s$-channel production $pp \xrightarrow{Z_\text{B}} jj$ leading to a dijet resonance, and the production in association with the baryonic Higgs $pp\xrightarrow{Z_\text{B}} Z_\text{B}S_\text{B}$.  Here we discuss the resonant $Z_\text{B}$ production with which we place a limit on the mass $m_{Z_\text{B}}$ and coupling $\alpha_\text{B}$ based on the absence of a $Z_\text{B}$ signal at the LHC, and leave the discussion of associated production to the next subsection.

Because dark matter relic abundance considerations suggest $m_\chi \gtrsim m_{Z_\text{B}}/2$ (see Fig. \ref{fig: DMPlots}), the decay channels to leptobaryon pairs 
\begin{equation}
Z_\text{B}\rightarrow \bar\psi\psi, \Sigma\Sigma, \chi\chi
\end{equation}
are closed.  This leaves the decay into standard model quark pairs as the dominant modes\footnote{Subdominant modes induced by fermion loops to final state  Higgs bosons $H$ and $S_\text{B}$, and to electroweak gauge bosons $W$, $Z$, and $\gamma$ are possible.  A subset of these are studied in \cite{Duerr:2015xxx} in the context of indirect detection in  model VA.}.  The gauge boson $Z_\text{B}$ universally couples to all flavors of quarks with equal strength, implying equality among the branching fractions to quark pairs.  
Therefore, the appropriate search strategy at the LHC is to look for a dijet formed by the lighter quarks ($u$, $d$, $c$, $s$, $b$) in the final state, along with a corresponding absence of a dilepton resonance to verify the leptophobic nature of $Z_\text{B}$.

Assuming massless quarks, the leading order parton-level cross section for a given flavor of initial and final state quarks is
\begin{align}\nonumber
\hat\sigma
& =  
\frac{4\pi\alpha_\text{B}^2}{243}\frac{\hat{s}}{(\hat{s}-m_{Z_\text{B}}^2)^2+m_{Z_\text{B}}^2\Gamma_{Z_\text{B}}} \\
& \approx \frac{4\pi^2 \alpha_\text{B}^2}{243}\frac{m_{Z_\text{B}}}{\Gamma_{Z_\text{B}}}\delta(\hat{s}-m_{Z_\text{B}}^2)\,, \qquad \text{({\sc nwa})}
\end{align}
where $\sqrt{\hat{s}}$ is the partonic center-of-mass energy and $\Gamma_{Z_\text{B}} = 2\alpha_\text{B} m_{Z_\text{B}}/3$ is the $Z_\text{B}$ total width.  In the second line, the narrow width approximation is made and is justified because $\Gamma_{Z_\text{B}}, m_q \ll m_{Z_\text{B}}$ \cite{Berdine:2007uv}.  The partonic cross section is convoluted with the leading order MSTW 2008 \cite{Martin:2009iq} parton distribution functions for $\sqrt{s} = 8$ TeV, and summed over the five light quark flavors yielding an estimate of the $pp\rightarrow jj$ cross section.  

The most stringent bound on the acceptance times cross section in the range $1\text{ TeV} \lesssim m_{Z_\text{B}} \lesssim 3\text{ TeV}$ comes from the recent CMS search for a dijet resonance \cite{Khachatryan:2015sja} with 19.7 $\text{fb}^{-1}$ of data.  Using their suggested acceptance factor of 0.6, we convert their limits into upper limits on $\alpha_\text{B}$ as a function of $Z_\text{B}$ mass, shown in Fig. \ref{fig:CMSDijetExclusion}.  Also shown in dashed lines are the more competitive limits in the lowest mass region \cite{Dobrescu:2013cmh} derived from the CDF \mbox{(0.11 $\text{fb}^{-1}$)} \cite{Abe:1997hm} data set and an older CMS analysis (4 fb$^{-1}$) \cite{Chatrchyan:2013qha}.

We find that constraints on the coupling constant depend strongly on the $Z_\text{B}$ mass.  Very roughly, the coupling is constrained to lie below $\alpha_\text{B}\lesssim 0.02$, except for the narrow mass range $1.2\text{ TeV} \lesssim m_{Z_\text{B}} \lesssim 1.7\text{ TeV}$ where the upper bound is at $\alpha_\text{B}\sim 0.008$.  The dark matter parameter scan of the previous section is indicated by the dashed green rectangle, and lies just below the $\alpha_\text{B}$ upper limit set by the recent CMS analysis.

\subsection{Decay and production of baryonic Higgs $S_\text{B}$}
\label{sec: SBProductionDecay}
As long as they are open, the largest decay modes of the baryonic Higgs $S_\text{B}$ for small $\theta$ are to a pair of leptobaryons or to a $Z_B$ pair
\begin{equation*}
S_\text{B} \rightarrow \psi\psi,\,\Sigma\Sigma,\,\chi\chi,\,Z_\text{B} Z_\text{B} \,.
\end{equation*}
Since the requirement of dark matter relic abundance saturation favors heavy dark matter in the non-resonant regime ($m_\chi \gtrsim m_{S_\text{B}}/2$), we expect the leptobaryon decay modes to be closed.  This leaves the $Z_\text{B}$ pair as the single dominant decay channel for small mixing angles $\theta\lesssim0.1$.  For larger mixing angles, decay modes to the $ZZ$, $WW$ and $HH$ bosons are dominant.

\begin{figure}
	\centering
	\includegraphics[width=65.0mm]{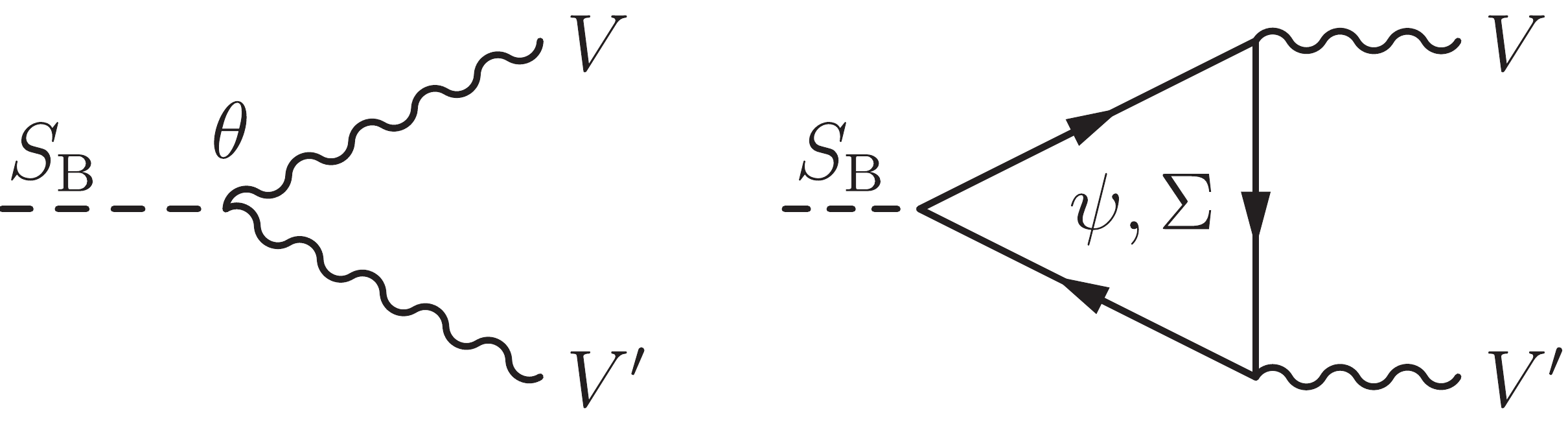}
	\caption{Representative Feynman graphs contributing to baryonic Higgs decay to electroweak gauge bosons $V,V' = \{W^\pm,\,Z,\,\gamma\}$.  The tree-level graph on the left is mixing angle suppressed compared to the loop diagram on the right.}
	\label{fig:higgsDecayFeynmanGraph}
\end{figure}

\begin{figure}
\includegraphics[width=\columnwidth]{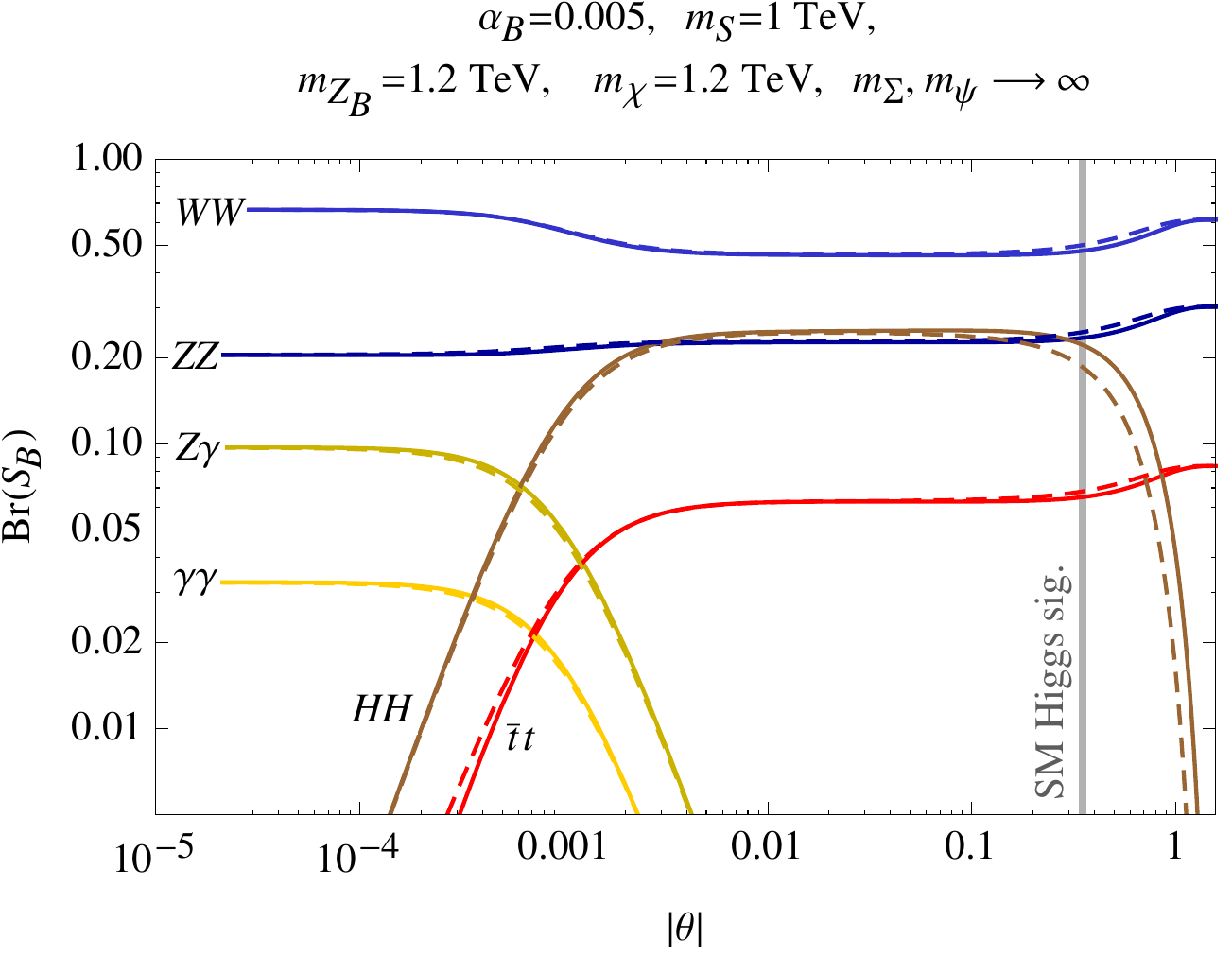}
\caption{Branching fraction of the baryonic Higgs boson $S_\text{B}$ as a function of the mixing angle $\theta$ [\mbox{$\theta>0$} solid, \mbox{$\theta<0$} dashed] below the $S_\text{B}\rightarrow Z_\text{B}Z_\text{B}$ threshold.  For mixing angles moderate in magnitude $|\theta|\gtrsim 10^{-3}$ standard model Higgs-like decay modes dominate, and for smaller mixing angles $|\theta| \lesssim 10^{-3}$ loop-mediated modes dominate (see Fig. \ref{fig:higgsDecayFeynmanGraph}).  The vertical line at $|\theta| = 0.35$ is the upper limit determined from Higgs signal strength measurements.}\label{fig:HiggsDecays}
\end{figure}

Below the $Z_\text{B}Z_\text{B}$ threshold, the baryonic Higgs decay profile exhibits a striking interplay between the mixing angle suppressed and the loop suppressed decay modes.  Particularly interesting are decays to electroweak gauge bosons shown in Fig. \ref{fig:higgsDecayFeynmanGraph}. The dramatic behavior of the branching fractions of $S_\text{B}$ as a function of the mixing angle $\theta$ is illustrated in Fig. \ref{fig:HiggsDecays} and is fairly independent of the particular values taken for the masses $m_{S}$, $m_{Z_\text{B}}$, $m_\chi$, coupling $\alpha_\text{B}$ and sign of the mixing angle.  For simplicity, masses of the charged leptobaryons have been taken to infinity (soft $S_\text{B}$ limit); we discuss the impact of finite leptobaryon masses consistent with vacuum stability shortly below.

For mixing angles moderate in magnitude $|\theta| > 10^{-3}$, in addition to the $HH$ mode, the baryonic Higgs inherits the largest decays of the standard model Higgs, which are the tree-level modes $WW$, $ZZ$ and $\bar{t}t$.  As the mixing angle is parametrically decreased, the standard-model-like decay modes give way to the leptobaryon loop-mediated decay modes.  In the small mixing angle regime $|\theta| < 10^{-3}$, the modes to electroweak gauge bosons dominate the total width.  The loop-mediated decay to a top quark pair is suppressed by $m_t^2/m_{Z_\text{B}}^2$ due to chiral symmetry.

The relative strengths of the decays to vector bosons in the limit $\theta\rightarrow 0$ are easily understood by integrating out the charged leptobaryonic multiplets $\Psi$ and $\Sigma$.  The effective Lagrangian describing the coupling of the baryonic Higgs field $s$ to the electroweak gauge bosons is
\begin{multline}\label{eq:effectiveLagrangian}
\mathcal{L_\text{eff}} =
\frac{g_\text{B}}{8\pi^2 m_{Z_\text{B}}} \sum_f\Big[e^2 Q_f^2 \,s F_{\mu\nu}F^{\mu\nu} \\
+ \frac{g^2}{c_\text{w}^2}(T_f^3 - Q_f s_\text{w}^2)^2 \, s Z_{\mu\nu}Z^{\mu\nu} \\
+ \frac{2 e g}{c_\text{w}} Q_f (T^3_f - Q_f s_\text{w}^2) \,s F_{\mu\nu}Z^{\mu\nu}\Big] \\
+ \frac{g_\text{B}}{8\pi^2 m_{Z_\text{B}}} \sum_F\Big[ g^2 c^{(W)}_F \,s W_{\mu\nu}^+ W^{- \mu\nu}\Big]\,,
\end{multline}
where the sum in the first three terms is over the isospin components $f=\{\psi^+,\psi^0,\Sigma^+,\Sigma^0\}$, and the sum in the final term runs over entire multiplets $F=\{\Psi,\Sigma\}$ where $c^{(W)}_\Psi = 1$ and $c^{(W)}_\Sigma = 2$.  The relative partial widths are obtained by squaring the couplings, and dividing the $ZZ$ and $\gamma\gamma$ modes by 2 to account for indistinguishability of final state particles.  This yields the relative branching ratios of the baryonic Higgs [see (\ref{eq:SBtoVVinmLBA})]
\begin{equation}\label{eq:SBDecayRatios}
\Gamma_{WW}:\Gamma_{ZZ}:\Gamma_{Z\gamma}:\Gamma_{\gamma\gamma} = 20 : 7 : 3 : 1\,.
\end{equation}
It is particularly interesting to compare this prediction with that of leptobaryon model VA.  The predicted ratios derived from the electroweak currents in Eq. (\ref{eq:extEWCurrentModelVA}) are
\begin{equation*}\tag{model VA}
\Gamma_{WW}:\Gamma_{ZZ}:\Gamma_{Z\gamma}:\Gamma_{\gamma\gamma} = 2 : 1 : 10^{-3} : 1\,.
\end{equation*}
The qualitative profile is substantially different from that of model A, with the $Z\gamma$ mode being far below the $\gamma\gamma$ mode in model VA.  Note that the precise numerical ratios are subject to mass corrections.  The largest corrections come from taking finite values of the charged leptobaryon masses to be consistent with the vacuum stability bound in Section \ref{sec:VacMetastab}.  However, the gross hierarchy of decay modes remains largely unchanged.  Therefore, for small mixing angles, we are optimistic about the prospects of distinguishing the leptobaryon models, even if the fermions cannot be directly produced at the LHC.\\

\begin{figure*}[t]
\includegraphics[width=\columnwidth]{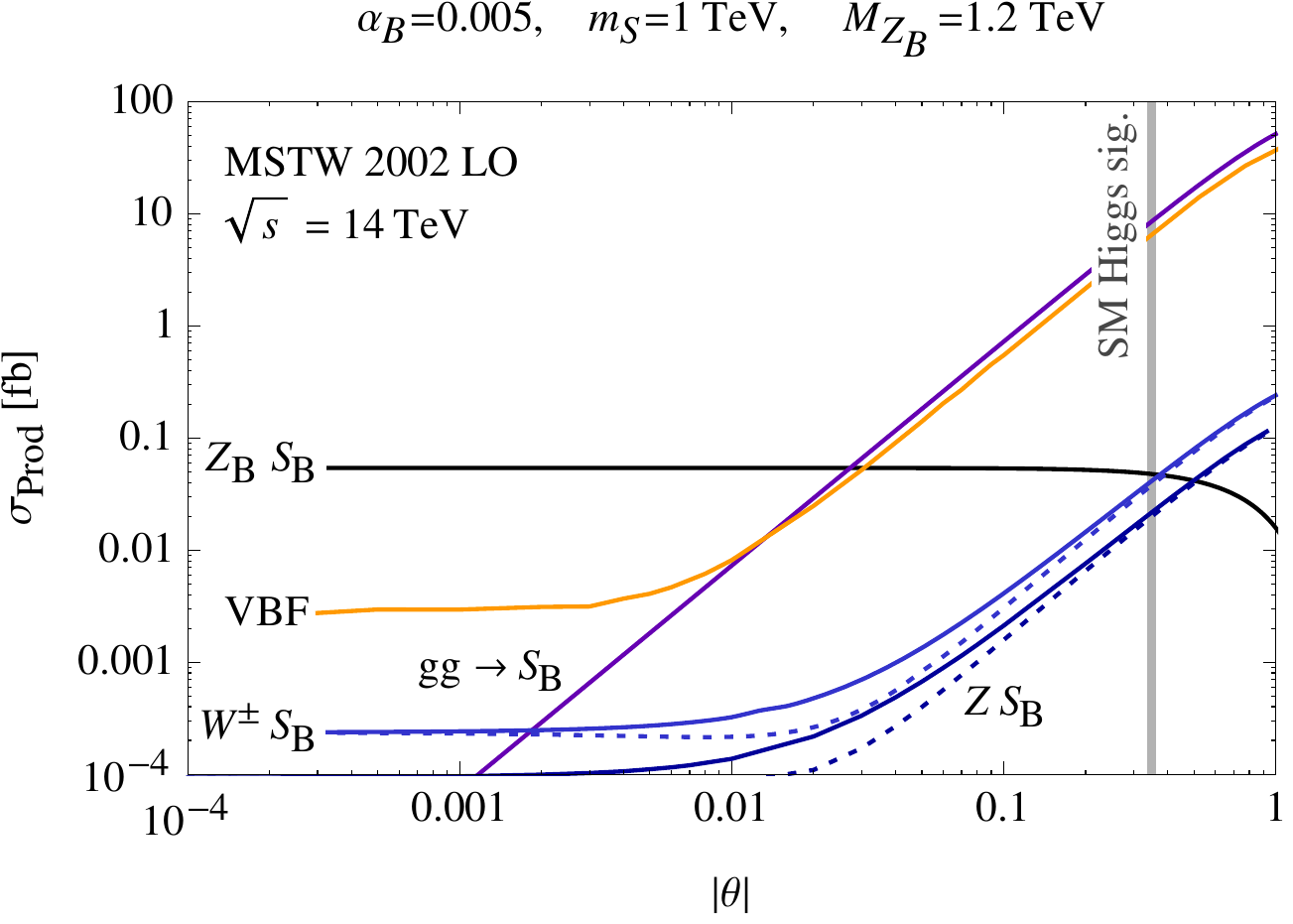}
\includegraphics[width=\columnwidth]{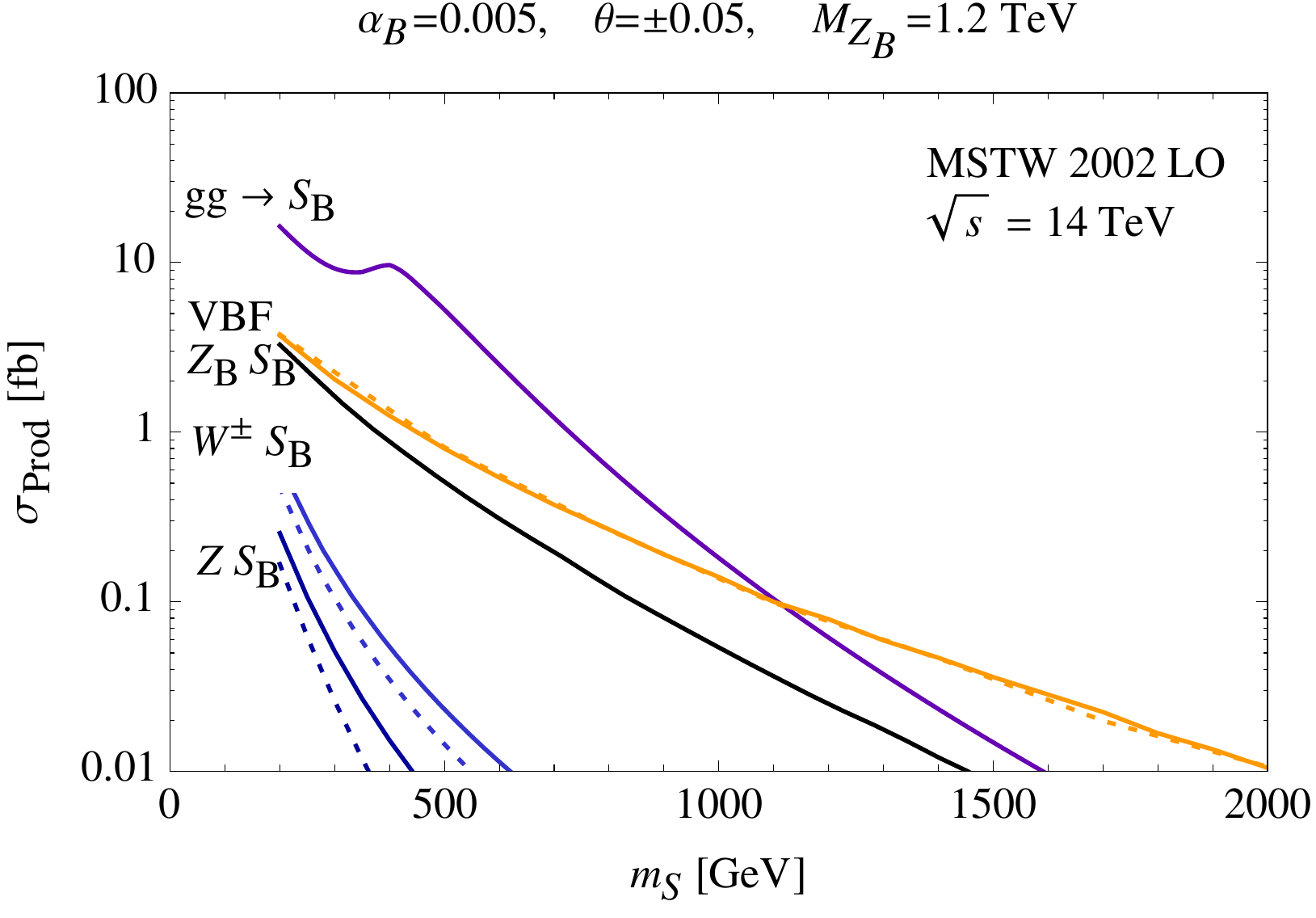}
\caption{Baryonic Higgs production cross sections at a $\sqrt{s} = 14\text{ TeV}$ $pp$ collider as a function of mixing angle (left) and $S_\text{B}$ mass (right) [\mbox{$\theta>0$} solid, \mbox{$\theta<0$} dashed].  The curve labeled ``VBF'' is the combined contributions of the mass suppressed $Z_\text{B}Z_\text{B}$, the mixing angle suppressed $WW/ZZ$, and the loop suppressed $WW$, $ZZ$, $Z\gamma$, and $\gamma\gamma$ fusion (see Table \ref{tab:SBProduction}).  The effect of charged leptobaryon loops is included by inserting the appropriate effective operator in Eq.~(\ref{eq:effectiveLagrangian}).  The vertical line at $|\theta| = 0.35$ is the upper limit determined from Higgs signal strength measurements.}\label{fig:HiggsProduction}
\end{figure*}

We turn to the production of the baryonic Higgs at the LHC.  Similar to its decay, the production cross sections depend on the mixing angle $\theta$ and also exhibit an interplay between mixing angle and loop suppressed contributions.  The production channels and representative parton-level Feynman graphs are listed in Table \ref{tab:SBProduction}.

Using the effective Lagrangian in Eq. (\ref{eq:effectiveLagrangian}) in the limit \mbox{$m_{\psi,\Sigma} \rightarrow \infty$}, we included the loop-mediated leptobaryon contributions to the vector boson fusion (VBF) and associated production channels.  In order to correctly handle interference and to facilitate the phase space integration, we obtained all production cross sections, except for gluon fusion, with {\sc CalcHEP 3.4} \cite{Belyaev:2012qa}, using the leading order MSTW 2002 parton distributions.  To keep finite top quark mass dependence, the one-loop gluon fusion cross section was computed separately with {\sc Package-X} \cite{Patel:2015tea}, and the results were convoluted with the MSTW 2008 PDF.  We checked our {\sc CalcHEP} implementation of leptobaryon model A by verifying that the tree-level predictions for $W^{\pm} S_\text{B}$ and $Z S_\text{B}$ production cross sections computed by hand and convoluted with the MSTW 2008 PDFs matched the {\sc CalcHEP} results.  This also confirmed that the results are fairly independent of the PDF data sets used.

The behavior of the cross sections is shown in Fig.~\ref{fig:HiggsProduction} for fixed $\alpha_\text{B}=0.005$ and $m_{Z_\text{B}} = 1.2$ TeV.  The panel on the left illustrates the mixing angle dependence of the various cross sections at fixed $m_S = 1$ TeV.  At small mixing angles, the tree-level $Z_\text{B}S_\text{B}$ cross section is the largest at 0.05 fb.  But, for larger mixing angles, the VBF and gluon fusion modes become larger.  The panel on the right shows the $m_S$ dependence at the mixing angle $|\theta| = 0.05$.  For light baryonic Higgs, the dominant mode of production is gluon fusion with a cross section reaching 20 fb.

\begin{table}[h!]
\begin{tabular}{l}
   \multicolumn{1}{l}{Associated production}\\[1mm]
	 \hline \multicolumn{1}{l}{$q\bar{q} \longrightarrow Z_\text{B}S_\text{B}$}\\[1mm]
	 $\parbox{2.5cm}{\includegraphics[height=19mm]{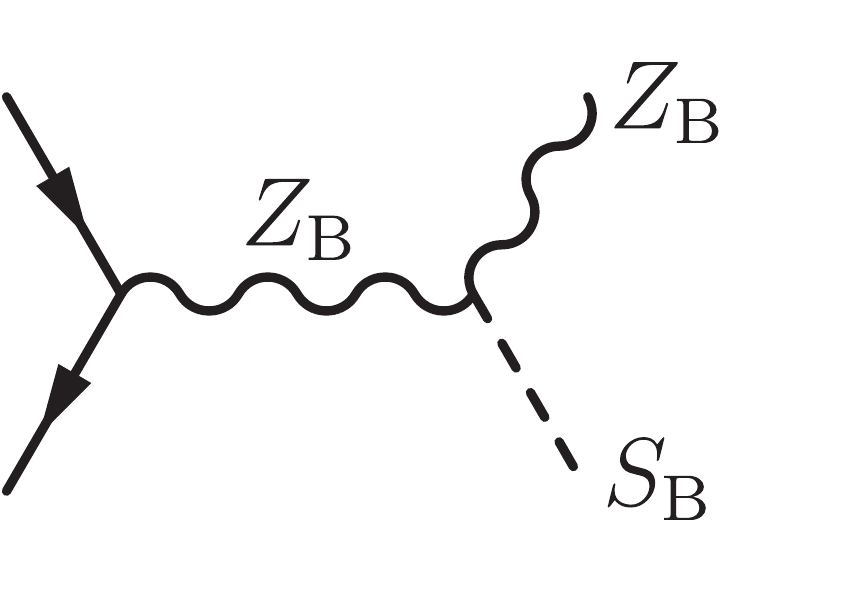}}$ \\ [5mm]
   
  \multicolumn{1}{l}{$q \bar{q} \longrightarrow W^\pm S_\text{B}, \enspace Z S_\text{B}$}\\
	\noindent $\parbox{2.5cm}{\includegraphics[height=19mm]{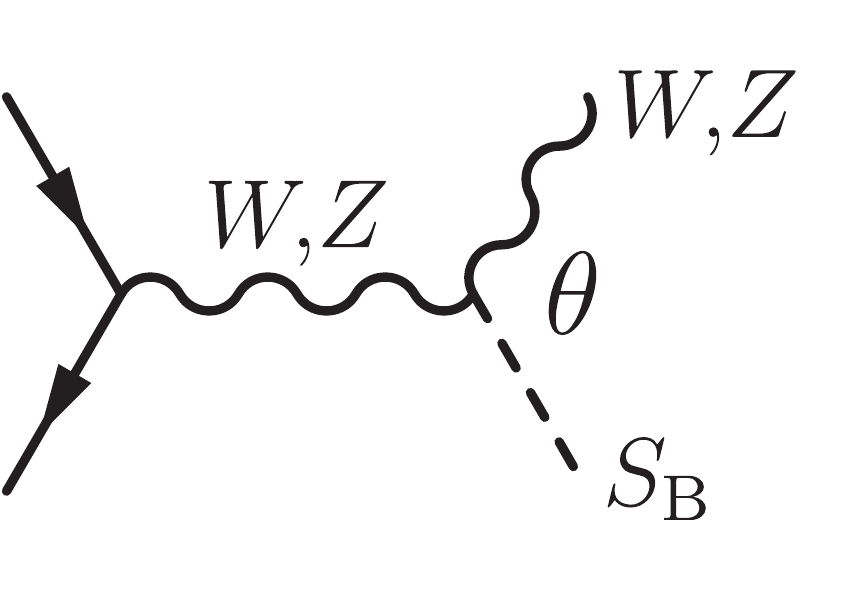}}  + 
	\hspace{3mm}\parbox{2.5cm}{\includegraphics[height=19mm]{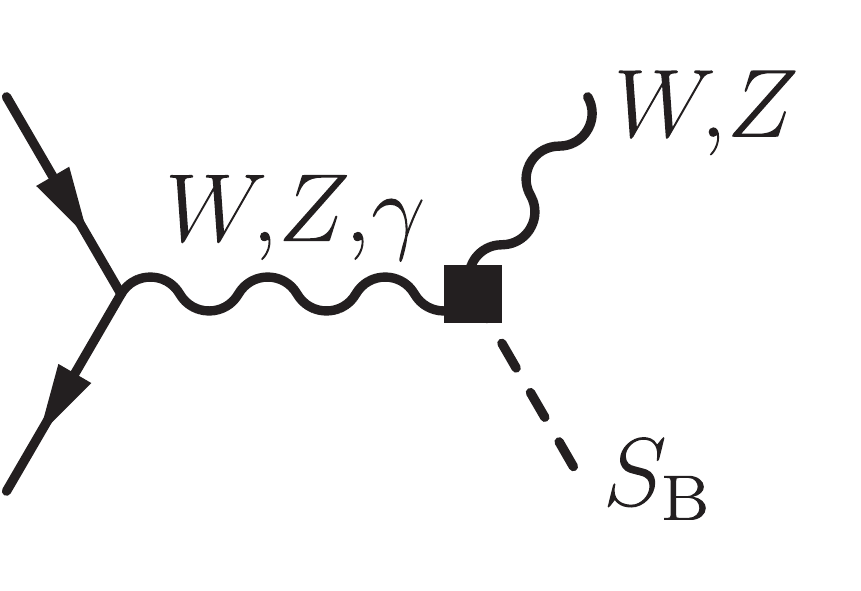}}$ \\[5mm]
  
	\multicolumn{1}{l}{Vector boson fusion (VBF)}\\
	\hline  \multicolumn{1}{l}{$qq \longrightarrow qq S_\text{B}$, \enspace$\bar{q}q \longrightarrow \bar{q}q S_\text{B}$, \enspace$\bar{q}\bar{q} \longrightarrow \bar{q}\bar{q} S_\text{B}$}\\[1mm]
	$\parbox{2.5cm}{\includegraphics[height=16mm]{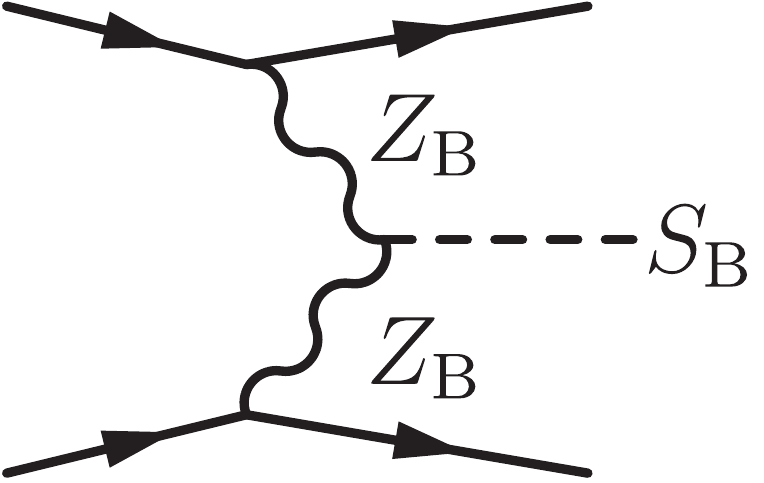}} \hspace{2mm}+ 
	 \hspace{-4mm}\parbox{2.5cm}{\includegraphics[height=16mm]{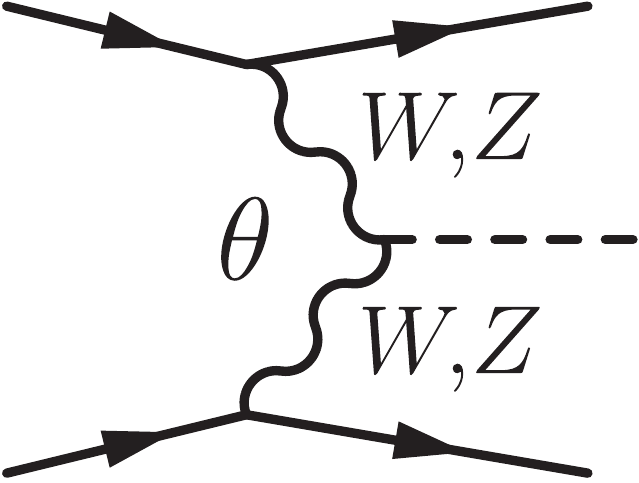}} + 
	 \hspace{-4mm}\parbox{2.5cm}{\includegraphics[height=16mm]{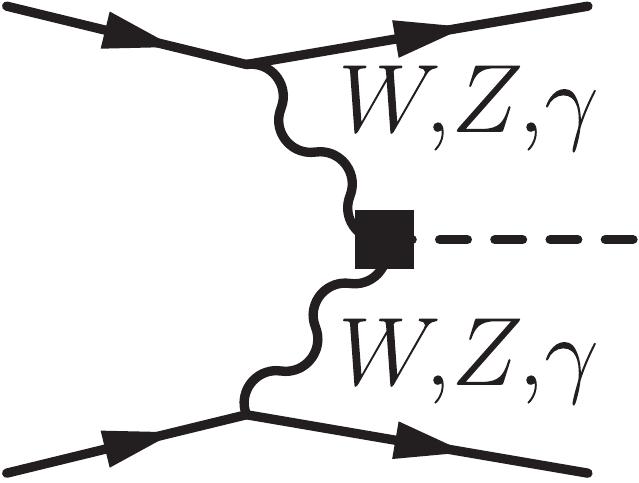}}$ \\[10mm]

  \multicolumn{1}{l}{Gluon fusion}\\
	\hline  \multicolumn{1}{l}{$gg \longrightarrow S_\text{B}$}\\[1mm]
	$\parbox{2.5cm}{\includegraphics[height=14mm]{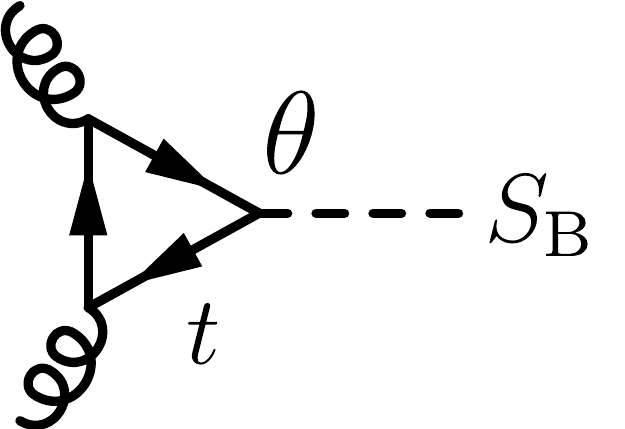}}$\\[7mm]
  \hline
\end{tabular}
\caption{Production channels of the baryonic Higgs $S_\text{B}$ at a hadron collider.  Vertices marked with a `$\theta$' are suppressed by the mixing angle, and square vertices are derived from the effective Lagrangian in Eq. (\ref{eq:effectiveLagrangian}). 
}
\label{tab:SBProduction}
\end{table}

The LHC search strategy for the baryonic Higgs is influenced by the production cross section and decay profile.  For mixing angles below $10^{-3}$, the $Z_\text{B}S_\text{B}$ production cross sections are quite small (0.05 fb), but the relatively clean $\gamma\gamma$, $Z\gamma$ and $ZZ$ decay modes are large and may make it possible to identify a signal at the LHC by searching for a resonance in the appropriate channel.  The current ATLAS limit on $\sigma_\text{fid}(S_\text{B})\cdot \text{BR}(S_\text{B}\rightarrow\gamma\gamma)$ \cite{Aad:2014ioa} at $\sqrt{s}=8$ TeV for the mass range 100--600 GeV is around \mbox{5 fb} (95\% C.L.), and is currently too large to place meaningful bounds.  For larger mixing angles  $|\theta| \gtrsim 0.02$ the gluon fusion mode is very large, although the only relatively clean decay mode is the $ZZ$.

However, by LHC run 3, the CMS and ATLAS experiments are expected to collect \mbox{1000--3000 $\text{fb}^{-1}$} of data at $\sqrt{s}=14$ TeV.  This will enable the experiments to probe the small mixing angle scenario for masses of $S_\text{B}$ and $Z_\text{B}$ reaching around 1 TeV (Fig. \ref{fig:HiggsProduction} left panel).  Therefore, we expect that the experiments at the LHC will be able to probe this model in the near future.

\section{Conclusions}
\label{sec: Conclusions}
In this article, we initiated an investigation of the dark matter and LHC phenomenology of low scale $\text{U(1)}_\text{B}$ symmetry breaking of leptobaryon model A.  
  Anomaly cancellation demands the existence of new colorless fermions $\psi$, $\Sigma$ and $\chi$ carrying baryon and lepton numbers, called \emph{leptobaryons}.  
The lightest leptobaryon $\chi$ is a Majorana fermion, and is automatically stabilized by an accidental $Z_2$ symmetry making it a dark matter candidate.
In this model, saturation of the observed dark matter relic abundance favors heavy dark matter ($m_\chi \gtrsim 1$ TeV), and null results by direct detection experiments lead to an upper limit on the Higgs mixing angle ($|\theta|\lesssim0.22$) that is competitive with current limits from LHC experiments ($|\theta| \lesssim 0.35$).  Future experiments will tighten these limits.  One-loop vacuum metastability places an upper limit on the leptobaryon masses not far above the baryonic gauge and Higgs boson masses.  Although it may be difficult to produce and study the leptobaryons at the LHC, it is possible to discover the $\text{U(1)}_\text{B}$ symmetry breaking mechanism by producing the baryonic gauge and Higgs bosons.  For small mixing angles, the branching fractions of the leptobaryon loop-mediated $S_\text{B}\rightarrow\gamma \gamma$, $Z\gamma$, $ZZ$ modes can reveal the underlying fermion content at the LHC.


Our initial study opens up new avenues for future investigations.   It may be interesting to speculate whether the other neutral leptobaryons $\psi^0$ or $\Sigma^0$ can populate the dark matter abundance.  Furthermore, in the limit of vanishing Higgs Yukawa couplings $\lambda_1 \ldots \lambda_4$ in Eq. (\ref{eq:YukawaWithH}) all neutral leptobaryons are stable, and give rise to multicomponent dark matter.   Within the context of electroweak baryogenesis, the addition of the baryonic Higgs leads to another scalar field participating in the electroweak phase transition.  Furthermore, the standard model Higgs coupling to the leptobaryons can accommodate new sources of $CP$ violation.  It may be worthwhile to investigate whether this model can account for the observed baryon asymmetry while evading constraints from low energy electric dipole moment measurements.  Finally, spontaneous symmetry breaking of U(1) gauge groups leads to the generation of Nielsen-Olson strings in the early Universe.  In light of the recent study by \cite{Long:2014mxa,Long:2014lxa}, it may also be interesting to investigate whether high scale $\text{U(1)}_\text{L}$ breaking can lead to observable signatures.  

\begin{acknowledgements}
The dark matter annihilation cross sections were computed with {\sc FeynCalc} \cite{Mertig:1990an}, and the loop-level partial widths were computed with {\sc Package-X} \cite{Patel:2015tea}.  The supplementary {\sc MadGraph 5} \cite{Alwall:2014hca} model file was generated with {\sc FeynRules 2.0} \cite{Alloul:2013bka}.  We thank Michael Duerr, Satoru Inoue, David Morrissey and Branimir Radovcic for helpful discussions while completing this work.  Special thanks goes to Pavel Fileviez P\'{e}rez for encouraging us to initiate this study and for providing ample feedback as this work was being completed.  We also thank Farinaldo Queiroz for assistance in preparing the Madgraph 5 model file.
\end{acknowledgements}

\appendix
\section{Leptobaryon model VA}
\label{sec: model VA}

\begin{table}
\begin{tabular}{ccccc}
\hline\hline
 & $\text{SU(2)}_L$ & $\text{U(1)}_Y$ & $\text{U(1)}_\text{B}$ & $\text{U(1)}_\text{L}$\\
 Gauge fields & $\vec{W}^\mu$ & $B^\mu$ & $Z_\text{B}^\mu$ & $Z_\text{L}^\mu$\\[1mm]
\hline Fermions\\

$\overline{\nu}_{\mathrlap{e,\mu,\tau}}$ & $\mathbf{1}$ & $0$ & $0$ & $-1$ \\
$\Psi$ & $\mathbf{2}$ & $-1/2$ & $\phantom{+}B_1$ & $\phantom{+}L_1$\\
$\overline{\Psi}$ & $\overline{\mathbf{2}}$ & $\phantom{+}1/2$ & $-B_2$ & $-L_2$\\
$\eta$ & $\mathbf{1}$ & $-1$ & $\phantom{+} B_2$ & $\phantom{+} L_2$\\
$\overline{\eta}$ & $\mathbf{1}$ & $-1$ & $- B_1$ & $-L_1$\\
$\chi$ & $\mathbf{1}$ & $\phantom{+} 0$ & $\phantom{+} B_2$ & $\phantom{+} L_2$\\
$\overline{\chi}$ & $\mathbf{1}$ & $\phantom{+} 0$ & $- B_1$ & $- L_1$
\\[1mm] \hline

Scalar fields\\
$H$ & $\mathbf{2}$ & $\phantom{+}1/2$ & $0$ & $0$ \\
$S_\text{B}$ & $\mathbf{1}$ & $0$ & $-3$ & $-3$ \\
$S_\text{L}$ & $\mathbf{1}$ & $0$ & $0$ & $2$ \\
\hline\hline
\end{tabular}
\caption{Field content of leptobaryon model VA, with the fermions given in two-component left-handed notation.  All fields are color singlets, and the hypercharge is normalized according to $Q = T^3 + Y$.}\label{tab:FieldContent-VA}
\end{table}
For reference, the formulation of leptobaryon model VA is briefly summarized and some salient phenomenological features of low scale $\text{U(1)}_\text{B}$ breaking are highlighted in this appendix.  

The gauge-Higgs sector is identical to that of model A.  But, in model VA six fermionic leptobaryon multiplets are added (see Table \ref{tab:FieldContent-VA}).    Upon spontaneous symmetry breaking of $\text{U(1)}_\text{B}$, the leptobaryons acquire vectorlike masses leading to three Dirac fermions: $(\psi^0, \psi^-)$, $\eta^-$ and $\chi$.  They couple to the standard model through the weak and electromagnetic vector currents given by
\begin{align}\label{eq:extEWCurrentModelVA}
\nonumber J_\text{EM}^\mu &= -\overline{\psi^-} \gamma^\mu \psi^- - \overline{\eta^-} \gamma^\mu \eta^-\\
\nonumber J_\text{NC}^\mu &= \textstyle \frac{1}{2} \overline{\psi^0}\gamma^\mu\psi^0 + (-\frac{1}{2}+s_\text{w}^2)\overline{\psi^-} \gamma^\mu \psi^- + s_\text{w}^2\overline{\eta^-}\gamma^\mu\eta^-\\
J_\text{CC}^{-\mu} &= \textstyle \frac{1}{\sqrt{2}} \overline{\psi^0}\gamma^\mu\psi^- \,,
\end{align}
and through the extended baryonic and leptonic currents, which have both vector and axial-vector parts
\begin{equation}\label{eq:extBaryonCurrentModelVA}
J_\text{B,ext}^\mu  = \textstyle J_\text{B,V}^\mu + J_\text{B,A}^\mu\,, \qquad
J_\text{L,ext}^\mu  = \textstyle J_\text{L,V}^\mu + J_\text{L,A}^\mu\,.
\end{equation}
The axial-vector parts are fixed by the anomaly cancellation condition ($B_1-B_2 = L_1-L_2 = -3$):
\begin{multline}
J_\text{B,A}^\mu = J_\text{L,A}^\mu = \textstyle \frac{3}{2} \big(\overline{\psi^0}\gamma^\mu\gamma_5\psi^0 + \overline{\psi^-}\gamma^\mu\gamma_5\psi^-\big) \\ 
- \textstyle\frac{3}{2}\overline{\eta^-}\gamma^\mu\gamma_5\eta^- -\frac{3}{2} \bar\chi\gamma^\mu\gamma_5\chi\,.
\end{multline}
But the overall normalization of the vector parts remains unconstrained
\begin{align*}
J_\text{B,V}^\mu  &=\textstyle B \big(\overline{\psi^0}\gamma^\mu\psi^0 + \overline{\psi^-}\gamma^\mu\psi^- +\textstyle \overline{\eta^-}\gamma^\mu\eta^- + \bar\chi\gamma^\mu \chi \big)\\
J_\text{L,V}^\mu  &=\textstyle L \big(\overline{\psi^0}\gamma^\mu\psi^0 + \overline{\psi^-}\gamma^\mu\psi^- +\textstyle \overline{\eta^-}\gamma^\mu\eta^- + \bar\chi\gamma^\mu \chi \big)\,,
\end{align*}
where $B\equiv B_1 + B_2$ and $L \equiv L_1 + L_2$.  

The fact that the leptobaryons have a vector part to their interactions with $Z_\text{B}$ qualitatively changes the dark matter phenomenology as compared to that of model A.  The direct detection cross section via $t$-channel $Z_\text{B}$ exchange is not velocity suppressed.  Also, the resonant annihilation channel $\chi\bar\chi \rightarrow Z_\text{B} \rightarrow q\bar{q}$ is not velocity suppressed.  This makes it possible  for lighter dark matter, and opens the possibility for signals in indirect detection experiments \cite{Duerr:2015xxx}.  Furthermore, the vector currents are associated with an accidental global ``$\eta$ symmetry'' that is anomaly free \cite{Perez:2013nra}, and leads to an interesting connection between the observed baryon asymmetry and the dark matter asymmetry \cite{Perez:2013tea}.

\section{Scalar potential parameter relations}
\label{sec: ScalarPotentialParameter}
For reference, we provide the tree-level relations between the scalar potential parameters in (\ref{eq:modelpotential}) and the masses, {\sc vev}s and mixing angle
\begin{align}\label{eq:potentialRelations}
\nonumber \mu^2 &= \textstyle \frac{1}{2}(m_H^2 \cos^2\theta + m_S^2 \sin^2\theta) \\
\nonumber & \textstyle \hspace{3cm}- \frac{v_\text{B}}{4v}(m_H^2-m_S^2)\sin(2\theta)\\
\nonumber \mu_S^2 &=  \textstyle\frac{1}{2}(m_H^2 \sin^2\theta + m_S^2 \cos^2\theta) \\
\nonumber &\hspace{3cm} \textstyle- \frac{v}{4v_\text{B}}(m_H^2-m_S^2)\sin(2\theta)\\
\nonumber \lambda &= \frac{1}{2v^2}(m_H^2 \cos^2\theta + m_S^2 \sin^2\theta)\\
\nonumber b &= \textstyle \frac{1}{2 v_\text{B}^2}(m_H^2 \sin^2\theta + m_S^2 \cos^2\theta)\\
 a &=  \textstyle \frac{1}{2v v_\text{B}}(m_S^2-m_H^2)\sin(2\theta) \,.
\end{align}
These formulas are valid for $|\theta| \leq \pi/4$.

\section{Baryonic gauge and Higgs boson partial widths}
\label{sec: Higgs-decay}
The partial widths of the baryonic gauge boson are given by\\[-6mm]
\begin{align}\label{eq:zbwidths}
\nonumber \Gamma(Z_\text{B} \rightarrow q \bar{q}) &= \frac{\alpha_\text{B}}{9}m_{Z_\text{B}}\Big(1+\frac{2 m_q^2}{m^2_{Z_\text{B}}}\Big)\sqrt{1-4\frac{m^2_q}{m_{Z_\text{B}}^2}}\\
\Gamma(Z_\text{B} \rightarrow \chi\chi) &= \frac{3\alpha_\text{B}}{8}m_{Z_\text{B}}\Big(1-4\frac{m^2_\chi}{m_{Z_\text{B}}^2}\Big)^{3/2}\,.
\end{align}

To avoid displaying the complicated interference terms in the $S_\text{B}$ partial widths, we give the formulas in two distinct limits: $\theta\rightarrow \pi/2$ and $\theta\rightarrow 0$.  In the limit \mbox{$\theta\rightarrow \pi/2$}, the tree-level standard model Higgs-like modes are dominating\\[-5mm]
\begin{align*}
\Gamma(S_\text{B} \to W W) &= \frac{G_F}{8\sqrt{2}\pi}m_S^3\Big(1-4\frac{m_W^2}{m_S^2}+12\frac{m_W^4}{m_S^4}\Big)\\
&\hspace{3.5cm}\times\sqrt{1-4\frac{m_W^2}{m_S^2}}  \nonumber \\
\Gamma(S_\text{B} \to Z Z) &= \frac{G_F}{16\sqrt{2}\pi}m_S^3\Big(1-4\frac{m_Z^2}{m_S^2}+12\frac{m_Z^4}{m_S^4}\Big)\\
&\hspace{3.5cm}\times\sqrt{1-4\frac{m_Z^2}{m_S^2}}  \nonumber \\
\Gamma(S_\text{B} \to \bar{t}t) &= \frac{3 G_F}{4\sqrt{2}\pi}m_t^2 m_S\Big(1-4\frac{m_t^2}{m_S^2}\Big)^{3/2} \,.
\end{align*}
In the limit $\theta\rightarrow 0$ the only tree-level mode is $S_\text{B} \rightarrow Z_\text{B} Z_\text{B}$.  The leptobaryon loop-mediated modes to electroweak gauge bosons are given in the limit $m_{W,Z,\gamma} \ll m_S \ll m_{\Psi,\Sigma}$, and can be derived from Eq. (\ref{eq:effectiveLagrangian})
\begin{multline*}
\Gamma(S_\text{B} \rightarrow Z_\text{B} Z_\text{B}) = \frac{9\alpha_\text{B}}{8}\frac{m_S^3}{m_{Z_\text{B}}^2}\Big(1-4\frac{m_{Z_\text{B}}^2}{m_S^2}+12\frac{m_{Z_\text{B}}^4}{m_S^4}\Big)\\
\times\sqrt{1-4\frac{m_{Z_\text{B}}^2}{m_S^2}}
\end{multline*}\\[-12mm]
\begin{gather}
\begin{aligned}\label{eq:SBtoVVinmLBA}
\Gamma(S_\text{B} \to W W) &= \frac{\alpha_\text{B} m_S^3}{32\pi^4 m_{Z_\text{B}}^2}\frac{9g^4}{4} \\
\Gamma(S_\text{B} \to Z Z) &= \frac{\alpha_\text{B} m_S^3}{32\pi^4 m_{Z_\text{B}}^2}\frac{g^4}{8c_\text{w}^4}(3-6s_\text{w}^2+4s_\text{w}^4)^2\\
\Gamma(S_\text{B} \to Z \gamma) &= \frac{\alpha_\text{B} m_S^3}{32\pi^4 m_{Z_\text{B}}^2}\frac{e^2g^2}{4}(3-4s_\text{w})^2\,\\
\Gamma(S_\text{B} \to \gamma \gamma) &= \frac{\alpha_\text{B} m_S^3}{32\pi^4 m_{Z_\text{B}}^2} 2e^4\,.
\end{aligned}
\end{gather}
The decay mode to $HH$ vanishes in both limits.  Here we give the tree-level partial width that is valid for all values of $\theta$\\[-5mm]
\begin{equation}
\Gamma(S_\text{B} \to HH) = \frac{c_{SHH}^2}{8\pi m_S}\sqrt{1-4\frac{m_H^2}{m_S^2}} \,,
\end{equation}
where\\[-5mm]
\begin{equation}
c_{SHH}=\frac{(2m_H^2+m_S^2)}{4}\sin(2\theta)\Big(\frac{\cos\theta}{v}+\frac{\sin\theta}{v_\text{B}}\Big) \,.
\end{equation}

\bibliography{mLB-SM}

\end{document}